# Symbolic Predicate-Guided Language Agents for Inverse Design of Perovskite Oxides

*Dong Hyeon Mok[1], Seoin Back[2,3,4,*], Victor Fung[5,*] and Guoxiang Hu[6,7,*]*

[1]Department of Chemical and Biomolecular Engineering, Institute of Emergent Materials, Sogang University, Seoul 04107, Republic of Korea

[2]KU-KIST Graduate School of Converging Science and Technology, Korea University, Seoul 02841, Republic of Korea

[3]Department of Integrative Energy Engineering, Korea University, Seoul 02841, Republic of Korea

[4]Institute for Multiscale Matter and Systems (IMMS), Ewha Womans University, Seoul 03760, Republic of Korea

[5] School of Computational Science and Engineering, Georgia Institute of Technology, Atlanta, GA 30332, USA.

[6] School of Materials Science and Engineering, Georgia Institute of Technology, Atlanta, GA 30332, USA.

[7] School of Chemistry and Biochemistry, Georgia Institute of Technology, Atlanta, GA 30332, USA.

AUTHOR INFORMATION

**Corresponding Author**

E-mail: sback@korea.ac.kr, victorfung@gatech.edu, emma.hu@mse.gatech.edu

**Abstract**

Efficient discovery of high-performance materials has been pursued through a variety of data- and AI-driven strategies, among which inverse design, generating materials from desired target properties, has emerged as an important paradigm. Large language models (LLMs) offer a complementary route for inverse materials design because their reasoning and in-context learning capability can be used not only to propose candidates but also to demonstrate interpretable design principles. In this work, we introduce a domain specific language (DSL)-guided strategy to improve the reasoning and design capability of LLM agents by translating natural language design rules into symbolic predicates encoded in a predefined chemistry DSL. These predicates allow the LLM agent to obtain statistical evidence from the accumulated materials data, enabling the agent to evaluate and refine its own reasoning during the design loop. Based on this strategy, we developed a multi-agent materials design framework, called Operational Rule-grounded CHEmical Search Through Reasoning Agents (ORCHESTRA), and applied it to the inverse design of double perovskite oxides under multiple target-property objectives. The results show that symbolic predicates help the LLM identify unsupported rules, validate newly proposed rules and improve the rule store over iterative design cycles. Compared with a strategy relying only on natural language rules, the DSL-guided framework showed the potential to improve materials design performance, particularly for challenging target properties. These findings suggest that mathematical and statistical grounding can enhance the reasoning capability of LLM agents in materials science and that LLM-based inverse design can be performed effectively without large task-specific datasets or additional model training.

## 1. Introduction

Since the introduction of data and AI-driven strategies into materials science, substantial efforts have been devoted to efficiently identifying target materials from vast chemical spaces[1-7]. In particular, after the concept of inverse design was introduced into inorganic materials research, shifting the design paradigm from searching for desired properties within a predefined materials space to constructing materials from target properties, a variety of strategies have been proposed to accelerate this process[8-10]. Early approaches largely relied on high-throughput virtual screening, in which machine learning surrogate models for simulations were used to rapidly evaluate large chemical spaces[11, 12]. More recently, generative models that sample candidate materials from learned latent chemical space have become a major strategy for inverse materials design[13-15]. Although these approaches have been widely demonstrated to be effective for designing materials with desired properties, their performance often depends on the availability of sufficiently large and representative dataset for the target materials system. As a result, they are well suited for discovering unexplored candidates within relatively well-characterized chemical spaces, but remain limited when applied to poorly explored systems where extensive computational data must first be prepared[16].

One possible route to address this limitation is to leverage the reasoning capability of large language models (LLMs). The chemical knowledge and reasoning ability of LLMs have been demonstrated in various scientific tasks, and even when their internal knowledge is insufficient for specialized domains, it can be supplemented through data-retrieval strategies such as retrieval augmented generation (RAG)[17-19]. Recent studies on LLM-based materials design have therefore employed iterative loops and multi-agent frameworks, in which hypotheses are generated and evaluated through self-reflection, and refined through mutual critique among agents[20-23]. These frameworks have shown the potential to design materials with target properties while reducing the need for task-specific model training and large pre-existing datasets. In addition, because their decisions are expressed in natural language, LLM-based design agents can provide interpretable design hypotheses and potentially reveal new materials design principles[24].

However, the fundamental limitations of LLM-based reasoning cannot be ignored. Due to the probabilistic nature of LLM generation, hallucinations or unsupported claims can occur even when the output appears chemically plausible[25-28]. Moreover, improving reasoning through self-reflection or multi-agent critique still relies on the reasoning capacity of the underlying language model itself, making it difficult to fully overcome model intrinsic limitations. If incorrect reasoning is accumulated and repeatedly used during an iterative design loop, it may misguide candidate generation and even degrade design performance compared with more conventional algorithmic or ML-based strategies. Rather than relying solely on natural language reasoning, LLM-based materials design can benefit from external algorithmic or statistical tools that assist the reasoning process by evaluating, correcting and grounding LLM-generated design hypotheses that are converted into a form that can be explicitly tested through mathematical and statistical analysis[29, 30].

In this study, we adopted a strategy that translates natural-language design rules into a chemistry domain-specific language (DSL)[31-33]. In this strategy, LLM agent not only propose design rules in natural language but also generates symbolic predicates by selecting appropriate symbols and operators from a predefined DSL. These predicates can be applied directly to the accumulated materials dataset, providing statistical information such as the number of matched candidates, precision, and lift. This allows the framework to evaluate whether each rule is empirically supported and to determine the range of materials to which the rule applies. By converting LLM-generated reasoning into executable predicates, the framework enables natural-language design hypotheses to be grounded, tested, and refined using data.

Based on this strategy, we developed a multi-agent materials design framework called Operational Rule-grounded CHEmical Search Through Reasoning Agents (ORCHESTRA). The framework consists of a design agent that proposes candidate materials, an interpretation agent that extracts design rules and symbolic predicates from the evaluated data, and a curate agent that manages the rule store based on statistical evidence. Through iterative cycles of candidate generation, property evaluation, rule extraction, and rule curation, ORCHESTRA performs inverse design while continuously refining its rule store. We evaluate the framework on designing double perovskite oxides (DPs) across multiple target properties, including the band gap, formation energy, and oxygen evolution reaction (OER) overpotential as a representative example. The OER overpotential target provides a particularly challenging test case because it depends on multiple adsorption intermediates rather than a single directly optimized descriptor[34]. The results demonstrate that symbolic predicate grounding can improve LLM-guided materials design for such challenging targets, while also providing interpretable and statistically testable design principles.

## 2. Results

### 2.1 Framework

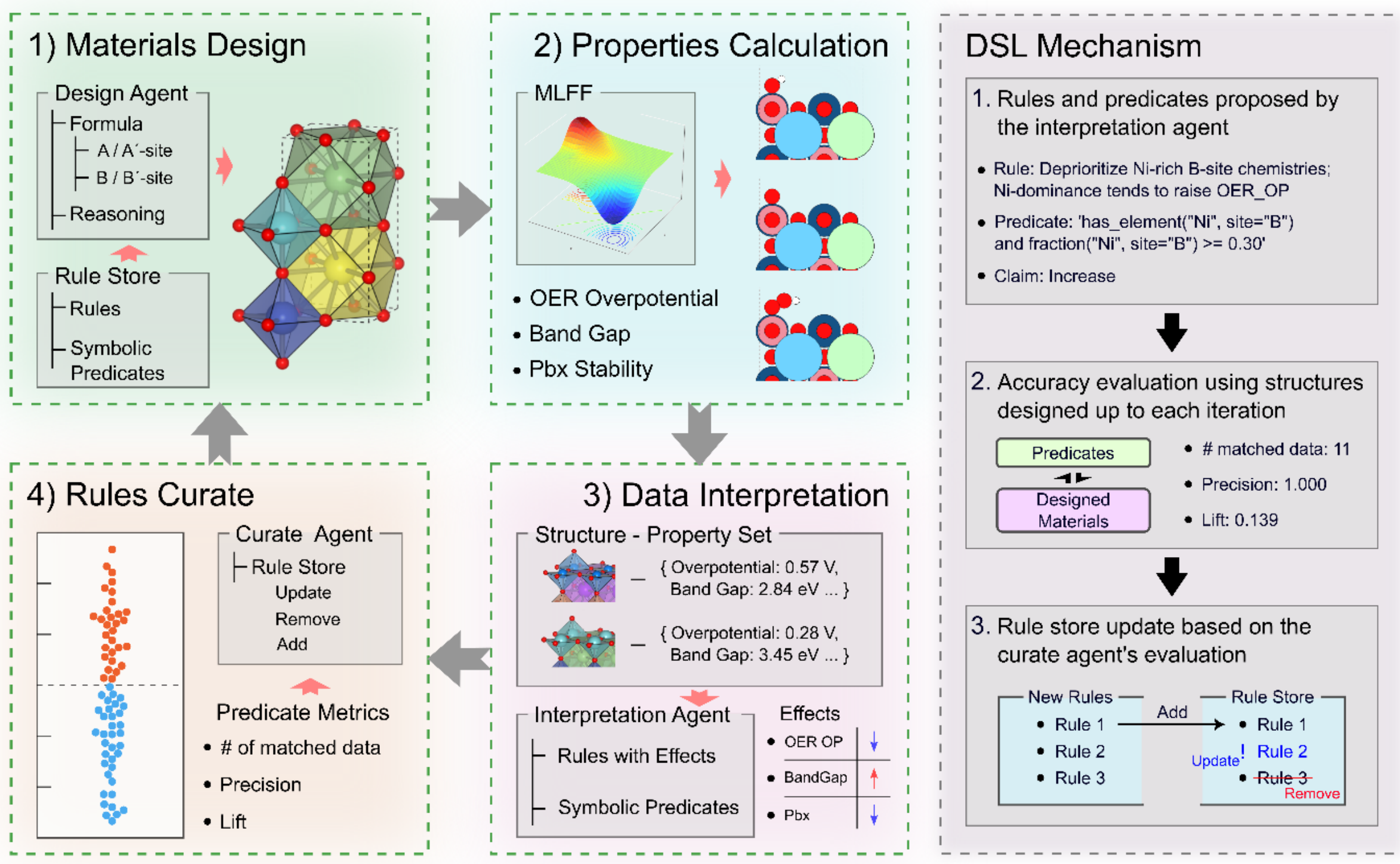

**Figure 1.** Overview of the ORCHESTRA framework and symbolic-predicate-based rule evaluation. The framework consists of four iterative stages. (1) The design agent proposes new material candidates using its chemical background knowledge, the accumulated design history, and the rules stored in the rule store. (2) Surrogate models optimize the generated structures and evaluate the target properties. (3) The interpretation agent analyzes the resulting structure property data and proposes natural language design rules together with executable symbolic predicates and target-specific effect directions. (4) The curate agent evaluates the proposed rules using predicate and adds, updates, or removes the rules before updating the rule store. The right panel illustrates how a symbolic predicate is applied to the accumulated dataset and how the resulting statistical evidence is used to update the rule store.

The proposed framework in this work is organized around four functional node types, each using a tool with a distinct role. The design node contains a design tool that proposes new material candidates based on any available interpretations and curated knowledge. The calculation node evaluates the target properties of given materials candidates using surrogate models. The interpretation node contains an interpretation tool that extracts intrinsic relationships between materials and their properties as design rules. The curate node validates the reliability of the rules, decides whether to store the proposed rules in the rule store, which can hold a certain number of rules, and provides these stored rules back to guide the design loops (**Figure 1**).

At the design node, an agent tasked with material design generates candidates based on its intrinsic chemistry background knowledge, the historical record of previously evaluated materials and their target values, and design principles stored in the rule store. At the interpretation node, the interpretation agent examined the set of designed materials and corresponding target properties to extract relationships underlying successful inverse design. In this work, we adopted an LLM-based interpretation agent. It analyzes the given dataset and proposes a bounded number of rules describing how compositional and structural features influence target properties, nominating compositions likely to achieve the desired values. These rules are forwarded to the curate agent for evaluation.

A major addition of the framework in this study is the symbolic predicate layer, implemented as a chemistry DSL that grounds each rule generated by the interpretation agent. Together with the natural language rule, the interpretation agent also emits a symbolic algorithmic counterpart, a Boolean-valued expression over a closed chemistry DSL, applicable to materials in the dataset. The predicate is evaluated on the materials accumulated up to that iteration, and produces data-driven statistics quantifying how accurately the rule's claim is empirically substantiated. At the curate node, together with the natural language rule, these statistics are then inspected by the curate agent that maintains the rule store. The rules with weak empirical support are revised or dropped, and the resulting store is capped at a maximum size so the rule set remains both accurate and concise. The complete system prompt templates used for each agent are provided in **Supplementary Note A**, and a detailed description of the symbolic predicate and DSL method is provided in **Supplementary Note B**.

The framework runs for a fixed number of iterations starting from a shared initial dataset of materials with pre-evaluated target properties. In this first iteration, no rules yet exist, so the design agent is conditioned only on the initial data and the LLM background knowledge. It proposes the pre-defined number of candidates partitioned into two roles, an exploration subset intended to broaden coverage of the chemical space rather than to hit the target, and an exploitation subset that explicitly targets the desired property values. Each candidate is then simulated with a machine learning force field (MLFF) to optimize the structures and obtain its target properties and is appended to the accumulated dataset. The updated dataset is then passed to the interpretation node, where the interpretation agent emits a bounded set of new rules with their symbolic predicates when applicable, and the curate agent integrates them with the existing rule store, which is in turn forwarded to the design node in the next iteration. After this iterative design, evaluation, interpretation and curation, the final outcome is assessed by the fraction or number of generated candidates satisfying the predefined target criteria for exploitation.

We compare four strategies that differ in the capabilities of the interpretation channel reaching the design agent. 1) Random, where no agent is involved, and candidates are sampled at random from the chemistry search space. 2) Background, where only the design agent runs, and the interpretation node is disabled. The design agent is conditioned on the running history of evaluated materials and its own chemistry background knowledge. 3) Rule-only, where the interpretation agent generates natural language rules but no symbolic predicates. Thus, the rules cannot be statistically validated within each iteration, and their evaluation instead relies on the curate agent's natural language judgment. The rule store consequently grows as a free-form knowledge base rather than a statistically verified rule store. 4) DSL-guided, where the full framework including symbolic predicates and statistical data verification of the rules.

### 2.2 Material System and MLFF Validation

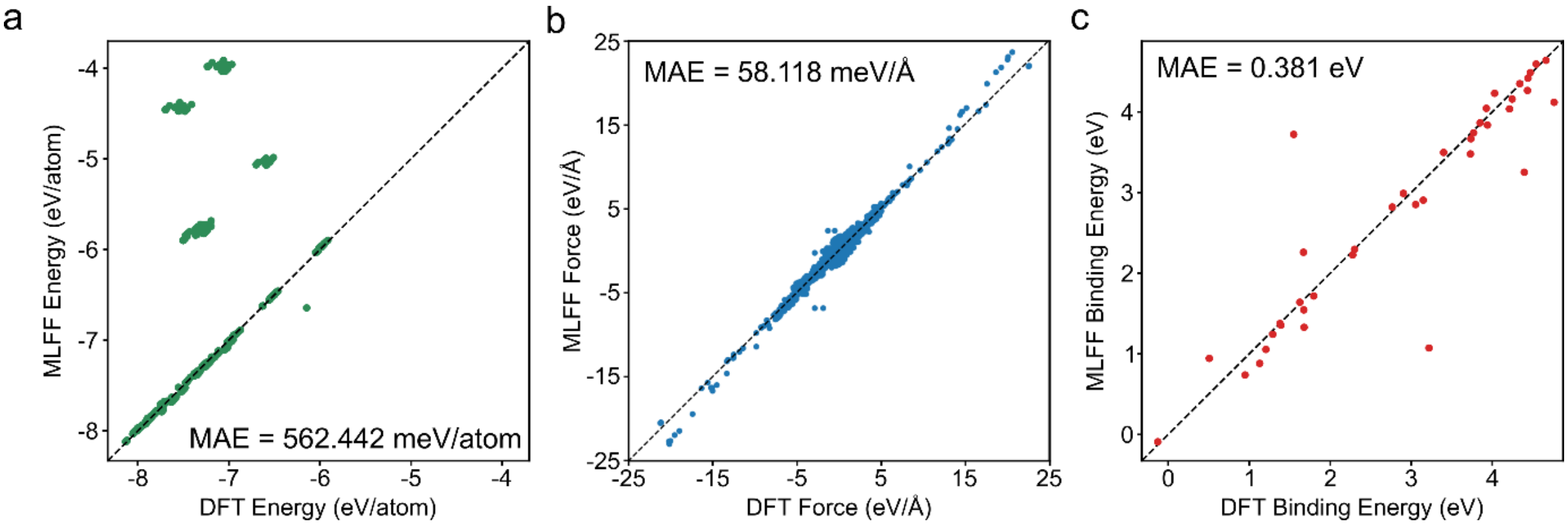

**Figure 2.** Parity plot between DFT-calculated and MLFF-predicted (a) energies, (b) forces and (c) binding Gibbs free energies of *O, *OH and *OOH on DP surface structures using UMA. The detailed values of DFT-calculated, MLFF-predicted and errors of binding Gibbs free energy can be found in **Table S7**.

To evaluate the materials design performance of the proposed framework, we selected oxide DPs with the formula $AA'BB'O_6$ as the target material system. Oxide DPs are promising for a broad range of applications, including electrocatalysis, fuel cells and photocatalysis[35, 36].

In addition, when the search space is restricted to cubic structures, structural complexity is reduced, allowing the agent to focus more directly on compositional information. Owing to the large number of possible elemental combinations in the $AA'BB'O_6$ space, this system provides a sufficiently broad yet controlled testbed for evaluating the proposed methodology (**Figure S1**)[37].

For efficient property evaluation within the framework, a MLFF capable of optimizing both bulk and surface oxide structures was required as a surrogate for density functional theory (DFT) calculations. In this study, we used the Universal Models for Atoms (UMA), specifically the uma-s-1p2 version, as the surrogate model[38]. Because UMA can be specialized for different materials through task-specific settings, the "OMat" task was used for bulk structure optimization of DPs, whereas the "OC22" task was used for surface structure optimization[39, 40].

Before adopting UMA as the surrogate model, we assessed whether it could provide a minimally reliable replacement for DFT in the present framework. For this validation, 14 DPs were randomly selected (**Table S7**), and their bulk and surface optimization performance, and binding Gibbs free energy prediction performance were compared with ground-truth DFT results. For bulk structures, UMA yielded mean absolute errors (MAEs) of 42.587 meV/atom for energy and 91.301 meV/Å for force (**Figure S2**). For surface structures, the corresponding MAEs were 562.442 meV/atom for energy and 58.118 meV/Å for force (**Figure 2a** and **b**). These results indicate that UMA did not achieve optimal surrogate model accuracy for either bulk or surface DPs. In particular, for several surface structures, the model failed to reproduce the DFT energy behavior.

Large errors were primarily observed for DPs containing certain lanthanide elements, such as $LaAlO_3$ and $SrLaMgNbO_6$, which may reflect the limited lanthanide-related training information available in the UMA model. Consistent with this interpretation, when La-containing perovskites were excluded from the validation set, the energy, force, and binding Gibbs free energy MAEs for surface structures improved substantially to 13.557 meV/atom 56.958 meV/Å and 0.141 eV, respectively (**Figure S3**). Even when La-containing perovskites were included, the Spearman rank correlation coefficient between the MLFF-predicted and DFT-calculated binding Gibbs free energies was 0.911, indicating strong agreement in their relative ordering. This result suggests that, despite systematic deviations in the absolute values, the MLFF largely preserves the DFT-derived binding Gibbs free energy trends across the evaluated materials.

The purpose of this validation was not to establish UMA as a quantitatively exact replacement for DFT, but rather to determine whether an available pretrained surrogate model could provide a sufficiently structured and chemically meaningful feedback signal for evaluating the operation of the proposed design framework. Although the absolute binding Gibbs free energy error of 0.381 eV and overpotential error of 0.370 V were non-negligible (**Figure S4**), the strong rank agreement supports the use of UMA as a computationally efficient oracle for comparative screening and trend-based materials design. Moreover, because all design strategies were evaluated using the same surrogate model, systematic MLFF errors were not expected to preferentially favor a particular strategy in the comparative framework evaluation. Accordingly, UMA was adopted as the surrogate model for the iterative design loop, and its predictions were interpreted as surrogate-defined estimates rather than quantitatively converged DFT values. The accuracies of other surrogate model candidates evaluated in this study are summarized in **Figure S5**.

## 2.3 Design Performance

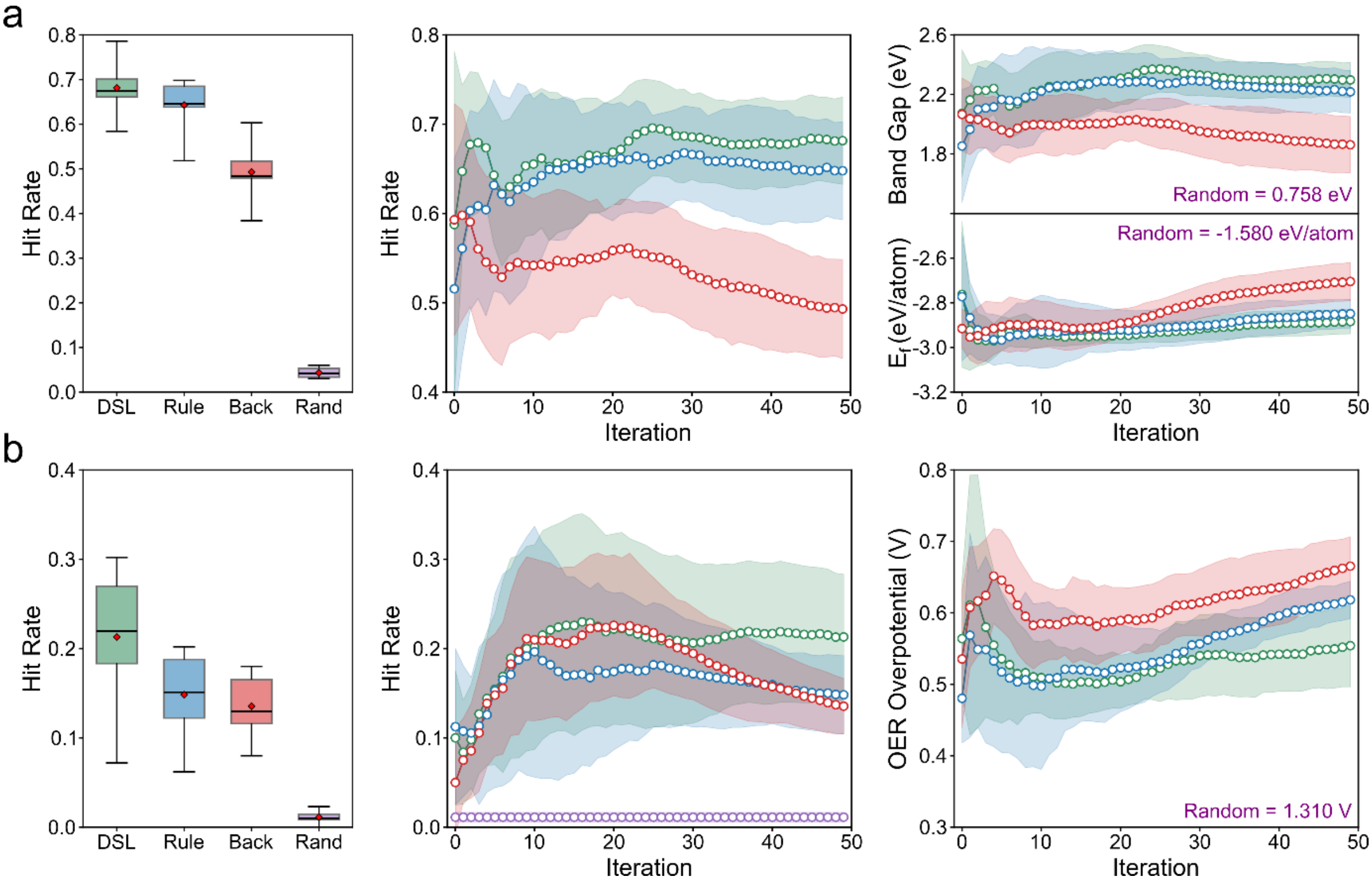

**Figure 3.** Inverse design performance of the four strategies. Results are shown for (a) the joint band gap/formation energy objective, defined by a band gap of at least 2.0 eV and a formation energy no greater than -2.0 eV/atom, and (b) the OER overpotential objective, defined by an overpotential of 0.3 V or lower. The left panels show the distribution of the final iteration hit rates across 10 independent runs. The middle panels show the mean hit rate as a function of design iteration, and the right panels show the corresponding mean target properties. Shaded regions represent the standard deviations across the 10 runs. Run-level hit rates are summarized in **Table S8.**

The material design performance of the framework was evaluated using the hit rate, defined as the fraction of non-duplicate materials generated for exploitation at each iteration that satisfied the predefined target criteria. To assess the robustness of the observed trends, all hit rates were averaged over 10 independent runs, each consisting of 50 design iterations. In addition to the hit rate, we separately monitored the generation rate, which represents the fraction of successfully generated non-duplicate materials at each iteration to the required number of generations for exploitation. This metric reflects the material generation efficiency of each strategy. Because a strategy that strongly biases the generated property distribution toward high-performing regions may also increase the probability of duplicate generation, the generation rate is expected to exhibit a partial trade-off with the hit rate (**Figure S6**). From an efficiency perspective, the absolute number of successful materials discovered over the same number of design iterations is also an important metric, as it directly reflects the total yield of target satisfying candidates produced by each strategy. These values are summarized in **Table S9**.

We evaluated the proposed DSL-guided strategy on two primary design tasks comprising a joint band gap/formation energy objective and an OER-overpotential objective. Band gap and formation energy were treated as a multi-target optimization problem, where the objective was to increase the band gap while decreasing the formation energy. A generated material was considered a hit when its band gap was at least 2.0 eV and, simultaneously, its formation energy was no greater than −2.0 eV/atom. For this target, the DSL-guided strategy achieved the highest hit rate, of approximately 0.681. The rule-only strategy without DSL also showed a similarly high hit rate, of 0.642, indicating that the band gap/formation energy multi-objective could also be effectively addressed using natural language design rules alone compared to strategy without any rules. By contrast, the lower hit rate of the Background strategy relative to the two rule-guided strategies suggests that explicitly providing design rules is more effective than relying on the design agent to interpret the accumulated data solely through its intrinsic chemical knowledge. Moreover, the Random strategy produced hit rates below 0.05, indirectly illustrating the difficulty of satisfying the predefined inverse design criteria through unguided sampling of the composition space. For the OER overpotential objective, DSL-guided strategy showed a clearer advantage. Although the overpotential is determined by the simultaneous optimization of three binding Gibbs free energies, $\Delta G_{*OH}$, $\Delta G_{*O}$ and $\Delta G_{*OOH}$, the design target was defined as a single objective, namely minimizing the OER overpotential[34]. A material was considered successful when its MLFF-calculated OER overpotential was 0.3 V or lower. Under this more demanding target, the DSL-guided strategy achieved a hit rate of approximately 0.213, the highest mean hit rate among the evaluated strategies.

We then analyzed how the statistical quality of the symbolic predicates used in the DSL-guided strategy evolved over the course of the design process. For this analysis, the precision and lift values of the symbolic predicates accumulated in the rule store were averaged over 10 independent runs and 50 design iterations. Precision measures the rate of matched materials that follow the claimed target direction, whereas lift quantifies how much this fraction exceeds the corresponding rate in the full dataset. Across the evaluated targets, precision and lift generally increased from their initial values during the early iterations and then stabilized, indicating that the rule store was progressively refined toward more reliable predicates (**Figure S7**). This trend is consistent with the dynamic update of the rule store, in which predicates with low statistical support were pruned or revised, whereas predicates with stronger evidence were retained or newly introduced. As a result, the average precision of the rules in the rule store gradually improved throughout the iterative design process. The corresponding computational time and token consumption are summarized in **Figure S8**.

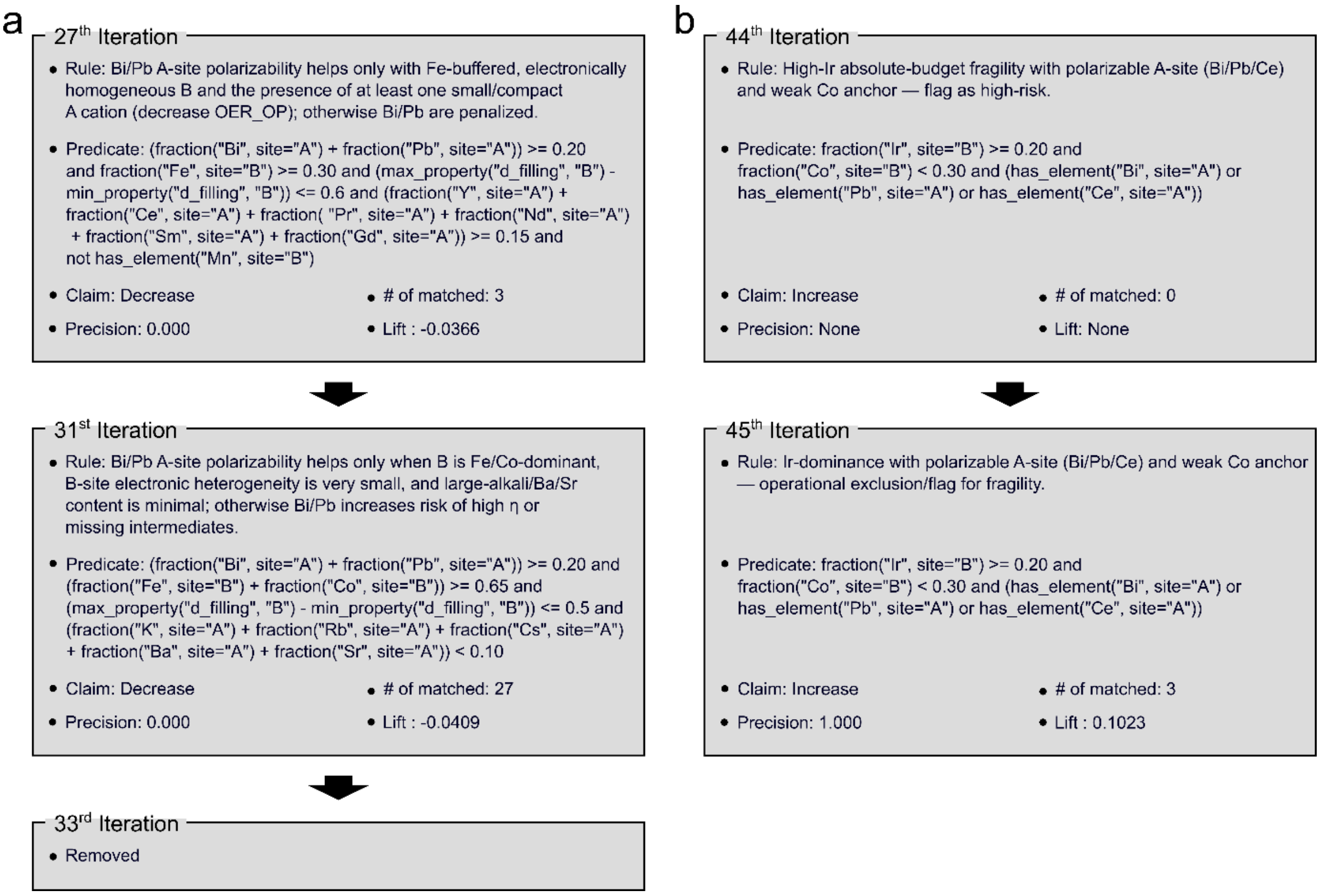

**Figure 4.** Representative examples of predicate guided rule refinement and prospective rule evaluation. (a) A Bi/Pb-related rule proposed at iteration 27 was repeatedly revised as its symbolic predicate continued to show zero precision and negative lift, and was ultimately removed from the rule store at iteration 33. (b) A rule proposed at iteration 44 initially had no matched materials and therefore could not be statistically evaluated. After additional materials entered the accumulated dataset, the same predicate matched three materials at iteration 45 and showed a precision of 1.00 and a positive lift of 0.102, providing preliminary empirical support for the proposed effect.

To further understand how symbolic predicates improved the design performance, we examined how individual rules evolved in the DSL-guided runs. Two representative advantages were observed. First, some rules initially exhibited low precision when evaluated through their symbolic predicates and were progressively modified before being removed from the rule store (**Figure 4a**). In a rule-only strategy, such rules could remain in the rule store because their validity cannot be explicitly adjudicated, potentially degrading subsequent design performance. As shown in **Figure S9**, unlike the DSL-guided strategy, in which both precision and lift increase, the rule-only strategy shows an increase in precision accompanied by a decrease in lift. This indicates that the rules generated by the rule-only strategy become biased toward rules associated with increased overpotential, while the overall reliability of the rules themselves decreases. In contrast, symbolic predicate-based verification allowed these unreliable rules to be quantitatively identified and excluded during the iterative process. Second, some rules could not be statistically evaluated at the time they were first proposed because no designed materials in the accumulated dataset matched their predicates. However, once new materials were generated based on these rules, the corresponding predicates immediately enabled data-driven validation of their accuracy (**Figure 4b**). This behavior indicates that the framework was not limited to interpolating rules from the initial dataset. Instead, it could propose rules that extrapolated beyond the currently verified data regime and then test those rules within subsequent iterations through the DSL. Such self-verifying extrapolative reasoning is not available in purely natural language rules, highlighting a key distinction between the proposed framework and the rule-only strategy.

The difference in performance across targets suggests that the benefit of symbolic predicate grounding depends on both the complexity of the target property and the degree to which the DSL can represent the relevant chemical constraints. The band gap/formation energy multi-target was comparatively easier to satisfy than the OER overpotential target. In this case, the hit rate exceeded 70% from the early stages of the design run and remained high until approximately the $25^{th}$ iteration. This behavior indicates that materials satisfying these relatively accessible criteria were discovered rapidly within the explored chemical space. As the remaining pool of undiscovered successful candidates became depleted, the hit rate of the DSL-guided strategy gradually decreased in later iterations. A more detailed discussion of the target-dependent hit rate is provided in **section 3.1**.

### 2.4 Analysis of Predicted OER Candidate

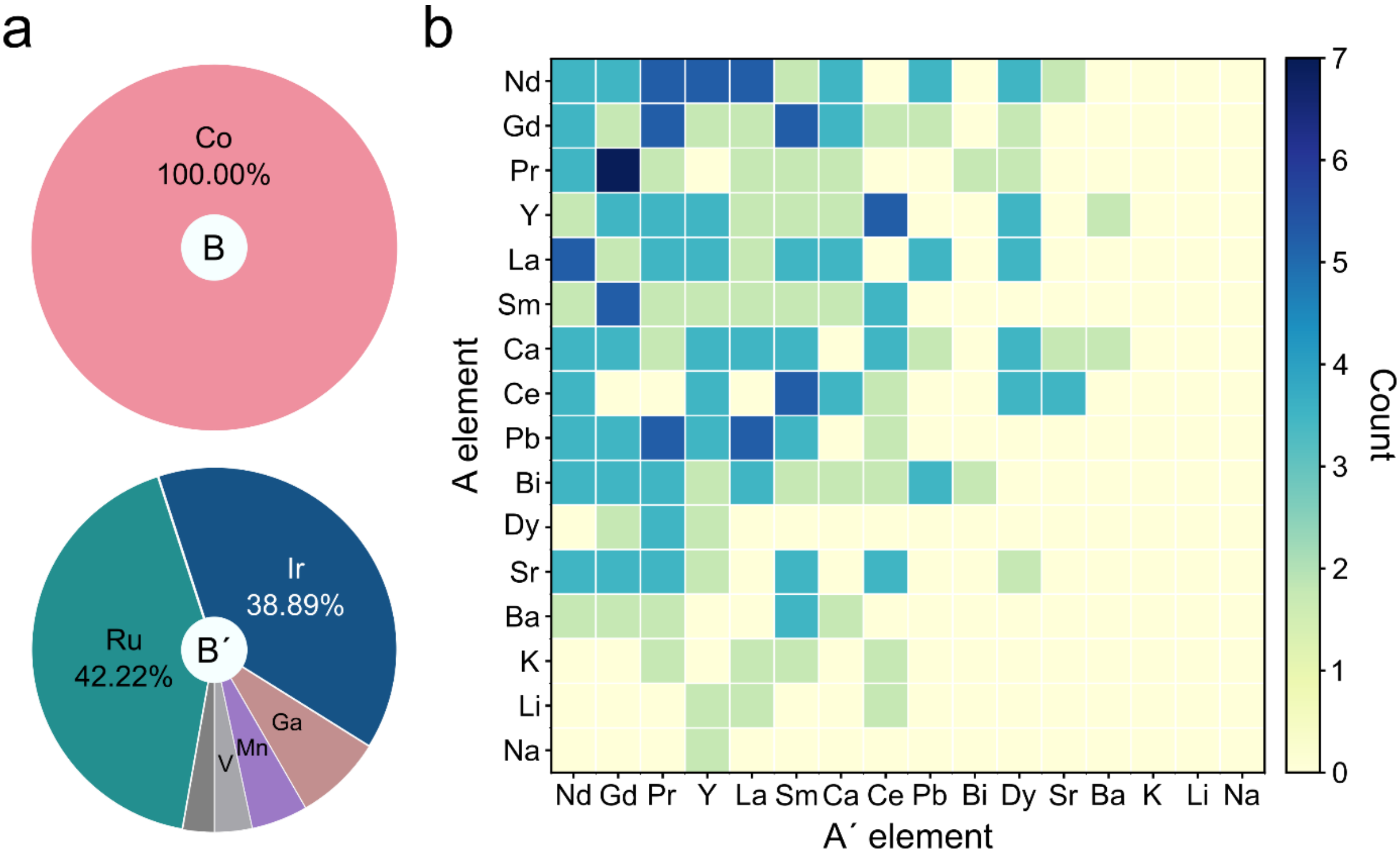

**Figure 5.** Compositional characteristics of DP candidates identified under the OER overpotential objective. The analysis includes 198 unique candidates with MLFF-predicted OER overpotentials of 0.3 V or lower. (a) Elemental distributions at the binding B site and the neighboring B′ site. (b) Co-occurrence matrix of A- and A′-site elements, where the color scale indicates the number of candidates containing each A/A′ combination.

The results above demonstrate that the proposed framework can achieve strong material design performance as an inverse design tool. Beyond this benchmark level evaluation, we further examined whether the framework could recover chemically plausible and literature-consistent compositional motifs from the MLFF-guided design runs. Across the 10 OER overpotential targeted runs, we collected the unique DPs whose MLFF-predicted OP was 0.3 V or lower. This procedure yielded a total of 198 unique DP candidates. A notable trend was that all of these candidates shared Co as the binding B-site element. In addition, more than 80% of the neighboring B′-site elements were Ru or Ir. This compositional motif is consistent with previous experimental observations that Co-Ir DPs exhibit high OER activity (**Figure 5**). The frequent occurrence of Nd, Pr and Sm at the A and A′-sites is also supported by experimental studies[41-43]. Importantly, this trend was not only observed in the generated candidates but was also reflected in the rules and symbolic predicates proposed by the agent. For example, one of the proposed rules was *'Co on B and (Ru or Ir) on B′, B-site mean d_filling in [5.0,6.2], limited A-site alkali (<0.30 total Li/Na/K/Rb/Cs), explicit A-site positive-charge buffering (Ca or moderate Pb/Bi or moderate lanthanide fraction), and no Ni on B -> lower OER_OP'*. This rule explicitly captures the beneficial roles of Co, Ru/Ir and lanthanide-containing A-site compositions in lowering the OER overpotential. In addition, several other rules and symbolic predicates also supported the high OER activity of Co-containing DPs (**Table S10**). These observations suggest that the framework extracted chemically meaningful and literature-consistent compositional principles during the iterative design process. However, because the individual generated candidates were not independently evaluated using higher-fidelity calculations or experiments, this candidate-level analysis should be interpreted as a qualitative assessment of chemical plausibility rather than as confirmation of their catalytic performance.

## 3. Discussion

### 3.1 Effect and Limitation of DSL by Target Property

Although the DSL-guided strategy achieved the best average hit rate for two primary design tasks including the joint band gap/formation energy task and the OER overpotential task, it did not consistently outperform the other strategies across all runs and target properties. In some runs, the DSL-guided strategy showed a hit rate substantially lower than its average performance, and for certain target properties, it even underperformed the rule-only strategy on average. To understand this behavior, we analyzed the relationship between the statistics of the symbolic predicates and the resulting design performance.

For the OER overpotential target, the average precision of the symbolic predicates across the 10 runs was not strongly correlated with the hit rate. More specifically, a high average precision did not necessarily guarantee a high hit rate, whereas runs with low predicate precision generally showed low hit rates. This behavior was closely related to the size of the chemical space covered by the symbolic predicates. For example, as shown in **Table S11**, runs 6 and 9 exhibited very high average precision, but the number of materials matched by their predicates was relatively small. This suggests that the predicates in these runs may have overfit to narrow chemical motifs, resulting in high precision without providing broadly generalizable design rules. In contrast, runs 2 and 4 covered a much larger portion of the dataset, but with very low predicate precision. These runs therefore appear to have generated overly broad rules that failed to capture meaningful trends in OER overpotential, leading to lower hit rates. These results indicate that successful design requires more than high predicate precision alone. Effective symbolic predicates must simultaneously achieve sufficient precision and cover an appropriate region of chemical space that is neither overly narrow nor overly broad, a balance that could be improved by introducing chemical space coverage metrics and instructing the curate agent to consider both statistical accuracy and applicability range.

Another limitation was observed for the $\Delta G_{*OH}$ single target optimization task, where the DSL-guided strategy showed a slightly lower average hit rate than the rule-only strategy (**Figure S10** and **S11**). To investigate this result, we compared the rule sets generated by the two strategies. First, the natural language rules generated by the rule-only strategy were substantially longer than those generated by the DSL-guided strategy. After 10 runs, the average length of the natural language rules stored in the rule store was 530 characters for the DSL-guided strategy and 1,218 characters for the rule-only strategy. Second, some rules generated by the rule-only strategy contained chemical descriptions that could not be expressed within the DSL defined in this study. Surface-related descriptions that were not explicitly supported by the DSL appeared, on average, 0.1 times per run in the DSL-guided strategy but 16.7 times in the rule-only strategy. Similarly, conditional 'if'-type statements outside the representational scope of the DSL appeared, on average, 0.6 times in the DSL-guided strategy and 10.1 times in the rule-only strategy (**Table S12**).

These observations suggest that requiring the interpretation agent to generate a symbolic predicate together with each natural language rule imposed an implicit constraint on the length and expressiveness of the resulting rules. When the most informative design rules for a given property involved concepts outside the predefined DSL, such as surface-specific motifs or complex conditional relationships, the DSL-guided strategy became less expressive than the rule-only strategy. As a result, the use of DSL predicates may reduce performance for targets whose governing chemical trends are not sufficiently covered by the current symbolic descriptor space. Nevertheless, this reflects the limitation of the current implementation of the DSL rather than an inherent weakness of symbolic design. It can be addressed by expanding the DSL to include a richer set of descriptors, and by decoupling the generation of natural language rules from that of their symbolic predicates. In such a framework, one agent could first propose chemically expressive natural language rules, while a separate formalization agent could subsequently translate only the representable components into executable predicates.

This would allow the framework to retain the flexibility of natural language reasoning while preserving the statistical grounding enabled by symbolic verification.

### 3.2 Effects of Framework and DSL Hyperparameters

The proposed framework contains several DSL-related components that can be adjusted as hyperparameters. We therefore analyzed how design performance was affected by factors such as the number of rules retained in the rule store, the number of exploitation candidates generated at each iteration, and the tolerance range used by the agent for the target property value. In particular, the current DSL includes predicates describing both the composition-related and elemental property-related features. Because most generated rules were observed to rely more frequently on predicates directly associated with elemental fractions than on comparative relationships between elemental properties (**Table S10** and **Figure S12**), we further evaluated the effect of reducing the DSL space by removing composition-related symbolic predicates from it.

As shown in **Figure S13**, the hit rate varied depending on hyperparameter settings. In particular, the strategy using the original DSL space achieved higher hit rates than the strategy in which composition-related predicates were excluded, indicating that the size and coverage of the DSL space are closely related to design performance. This result suggests that constructing a DSL appropriate for the target property is an important factor in improving the efficiency of the design process. However, we note that maximizing the hit rate is not necessarily equivalent to obtaining broadly generalizable design principles. As illustrated in **Table S10**, symbolic predicates that explicitly encode the presence of specific elements can sometimes lead to highly element-specific rules that risk overfitting to narrow chemical motifs. Therefore, when the objective is to extract more general design rules rather than to maximize immediate design performance, it may be advantageous to restrict such element-identity predicates and instead rely more heavily on predicates based on elemental properties.

## 4. Conclusion

In this study, we demonstrated inverse materials design from target properties by leveraging the reasoning capability of LLM agents, focusing on oxide DPs as a representative materials space. While existing materials design agents often rely on self-critique through chain-of-thought reasoning or cross evaluation among multiple agents to assess the reliability of their own reasoning, such approaches can become less effective as the context window grows and remain fundamentally constrained by the intrinsic reasoning capacity of the underlying language model. To address these limitations, we introduced a DSL-guided strategy that translates LLM-generated natural language chemical reasoning into symbolic predicates defined within a chemistry-specific DSL space. These predicates enable immediate, data-driven statistical evaluation of the proposed design rules.

Using this strategy, we found that the DSL-guided framework achieved higher design performance than a natural language rule-based LLM reasoning, particularly for complex targets such as OER overpotential in oxide DPs, which is difficult to optimize using conventional inverse design strategies. This improvement arises because the DSL-guided framework can explicitly evaluate the accuracy of generated reasoning against the accumulated dataset, remove unsupported or misleading rules, and test newly proposed hypotheses by generating candidate materials and verifying whether the resulting materials are consistent with the proposed predicates. Such rule-level validation is not available in a purely natural language rule-based strategy.

Furthermore, the design rules and DP candidates proposed by the framework for low OER overpotential reflected experimentally reported design principles and known high-performing compositional motifs. This result indicates that the framework can not only prioritize candidates satisfying defined target criteria but also extract chemically meaningful design principles from the iterative design process. Overall, our framework shows that high performance inverse materials design can be achieved through LLM agent reasoning without requiring system-specific training datasets or additional model training, while also providing a practical strategy for grounding and improving chemical reasoning through executable symbolic validation.

## 5. Methods

### 5.1 Structure Details

To evaluate the performance of ORCHESTRA, we selected oxide DPs as a representative materials system. The nominal composition is $AA'BB'O_6$, where the candidate elements available for each site are summarized in **Figure S1**. In this composition space, A and A′ are allowed to be the same element, and B and B′ are also allowed to be the same element. Therefore, the search space naturally included not only fully substituted $AA'BB'O_6$ compositions, but also simpler cases such as $ABO_3$, $AA'B_2O_6$ and $A_2BB'O_6$.

All bulk DP structures were constructed using a cubic perovskite-derived lattice. Excluding the oxygen sublattice, each bulk structure consisted of four cation layers, each containing four cation atoms. The A/A′ and B/B′ elements were arranged in an alternating pattern so that identical cation species were not directly connected through the oxygen lattice when two distinct elements occupied the same site type. Surface slab structures were generated from the corresponding bulk structures using the same four cation layer geometry. The slab was constructed such that a B-site containing layer was always exposed at the surface. A vacuum spacing of 10 Å was introduced on both sides of the slab, and the bottom two cation layers along with the oxygen atoms were fixed during geometry optimization. As a result, the exposed surface layer contained two B-site atoms and two B′-site atoms. For binding Gibbs free energy calculations, adsorbates were always placed on the B-site atoms.

### 5.2 Large Language Models

LLMs are autoregressive generative models based on the transformer architecture and are trained on large collections of text data[44]. Through this training process, they acquire statistical, semantic and contextual representations of language, enabling them to perform tasks involving reasoning, planning and decision-making in a human interpretable manner. In this work, all LLM-driven agents were implemented using OpenAI's GPT-5-mini model with the default temperature and top-p values, unless stated otherwise.

### 5.3 Surrogate Models

MLFF models provide efficient surrogates for DFT calculations by learning the relationship between atomic structure and potential energy surfaces from large scale quantum mechanical datasets[45]. Given the geometry and chemical composition of a material, these models predict total energies and atomic forces, enabling rapid geometry optimization and property evaluation at substantially reduced computational cost. MLFFs generally represent atomistic systems as graphs, where atoms are encoded as nodes with elemental features and interatomic interactions are represented as edges containing geometric information.

In this study, we employed UMA as the MLFF for oxide DP structures[38]. UMA is a family of universal machine learned interatomic potentials trained on a large collection of three-dimensional atomic structures spanning multiple chemical domains, including molecules, inorganic materials, and catalytic systems. The model is built on the equivariant Smooth Energy Network (eSEN)[46], which represents local atomic environments using spherical harmonic embeddings and propagates geometric information through equivariant message passing layers. UMA further incorporates a Mixture of Linear Experts (MoLE) architecture[47], which increases model capacity while maintaining efficient inference. Because UMA supports domain-specific prediction through task selection, we used the 'OMat' domain for bulk structures and the 'OC22' domain for surface structures[39, 40].

For band gap evaluation, the bulk structures were first optimized using UMA, and the band gaps of the MLFF-optimized structures were then predicted using a MEGNet surrogate model[48]. MEGNet is a graph network framework graph network framework for property prediction in molecules and crystals. It represents crystalline materials using graph-based inputs consisting of atomic, bond, and global state features, and has been shown to predict crystal properties such as formation energy, band gap and elastic moduli from the Materials Project database[49].

### 5.4 DFT Calculations

DFT calculations were performed using the Vienna Ab initio Simulation Package (VASP, version 5.4.4)[50] to pre-validate the MLFF performance. To maintain consistency with the reference data used for UMA, we adopted the same DFT settings as those used in the OC22 dataset. The projector augmented wave (PAW) method[51] was employed together with the Perdew-Burke-Ernzerhof generalized gradient approximation (GGA-PBE) exchange correlation functional[52]. All structures were fully relaxed until the total energy and maximum atomic force converged below $10^{-4}$ eV and 0.05 eV/Å, respectively. A plane-wave kinetic energy cutoff of 400 eV was used. Monkhorst-Pack k-point meshes[53] were set to $3 \times 3 \times 3$ for bulk DP structures and 3 × 3 × 1 for surface structures.

**5.5 Formation Energy, Gibbs Free Energy and Overpotential**

The formation energy of each oxide DP was calculated using elemental chemical potentials referenced from the Materials Project database. For a perovskite with the nominal composition AA′BB′O6, the total formation energy was computed as

$$E_f^{tot} = E^{AA'BB'O_6} - \mu_A - \mu_{A'} - \mu_B - \mu_{B'} - 6\mu_O$$

where $E^{AA'BB'O_6}$ is total energy of the relaxed bulk perovskite structure, $\mu_A$, $\mu_{A'}$, $\mu_B$, $\mu_{B'}$ and $\mu_O$ are the reference chemical potentials of A, A′, B, and B′ cation elements and oxygen, respectively. For compositions in which A = A′ or B = B′, the same expression was applied with the corresponding elemental chemical potential counted according to its stoichiometric multiplicity. The formation energy per atom ($E_f$) was obtained by dividing $E_f^{tot}$ by the total number of atoms in the formula unit.

The binding Gibbs free energies of oxygen evolution reaction (OER) intermediates on oxide DP surfaces were evaluated using the computational hydrogen electrode (CHE) approach[54], in which the chemical potential of a proton-electron pair ($H^+ + e^-$) is referenced to ½$H_2$ (g) at 0 V versus the reversible hydrogen electrode (RHE) under standard conditions. Based on this framework, the binding Gibbs free energies of *OH, *O and *OOH were calculated as follows:

$$\Delta G_{O*} = E_{O*} - E_{slab} - E_{H_2O} + E_{H_2} + \Delta(\text{ZPE} + \int C_p dT - TS)$$

$$\Delta G_{OH*} = E_{OH*} - E_{slab} - E_{H_2O} + \frac{1}{2}E_{H_2} + \Delta(\text{ZPE} + \int C_p dT - TS)$$

$$\Delta G_{OOH*} = E_{OOH*} - E_{slab} + 2E_{H_2O} + \frac{3}{2}E_{H_2} + \Delta(\text{ZPE} + \int C_p dT - TS)$$

Here, $E_{slab}$ denotes the total energy of the clean oxide DP surface, and $E_{*O}$, $E_{*OH}$ and $E_{*OOH}$ are the total energies of the surface with the corresponding adsorbed intermediates. $E_{H_2}$ and $E_{H_2O}$ represent the total energies of gas phase $H_2$ and $H_2O$, respectively. The correction term includes zero-point energies (ZPE), enthalpic ($\int C_p dT$), and entropic contribution terms (TS). These corrections were calculated using the Harmonic oscillator approximation for adsorbed intermediates and the Ideal gas approximations for gas molecules, as implemented in Atomic Simulation Environment (ASE)[55]. The correction values are given in **Table S13**.

The theoretical OER overpotential ($\eta_{OER}$) was determined from the four elementary proton-electron transfer steps of the conventional adsorbate evolution mechanism:

$$H_2O \rightarrow *OH + H^+ + e^-$$

$$*OH \rightarrow *O + H^+ + e^-$$

$$*O + H_2O \rightarrow *OOH + H^+ + e^-$$

$$*OOH \rightarrow O_2 + H^+ + e^-$$

At U = 0 V versus RHE, the corresponding free energy changes were defined as:

$$\Delta G_1 = \Delta G_{*OH}$$

$$\Delta G_2 = \Delta G_{*O} - \Delta G_{*OH}$$

$$\Delta G_3 = \Delta G_{*OOH} - \Delta G_{*O}$$

$$\Delta G_4 = 4.92 - \Delta G_{*OOH}$$

where 4.92 eV corresponds to the total equilibrium free energy for the four-electron water oxidation reaction. At an applied potential U, each proton-electron transfer step is shifted by −eU. Therefore, the OER limiting potential was calculated as:

$$U_{L,OER} = \max[\Delta G_1, \Delta G_2, \Delta G_3, \Delta G_4]/e$$

and the theoretical OER overpotential was obtained as:

$$\eta_{OER} = U_{L,OER} - 1.23$$

where 1.23 V is the equilibrium potential for water oxidation under standard conditions.

**Code Availability**

The code developed in this work and relevant information can be found in GitHub (https://github.com/ahrehd0506/Catalyst-Design-Agent)

**Acknowledgements**

D.H.M. acknowledges the support from Korea Institute for Advancement of Technology (KIAT)

grant funded by the Ministry of Trade, Industry & Energy (MOTIE), Korea Government (RS-2024-00436106, Human Resource Development Program for Industrial Innovation). S.B. acknowledges the support from the National Research Foundation of Korea (NRF) grants funded by the Korea government (MSIT and MOE) (RS-2024-00448287, RS-2025-16063688, RS-2025-00513832, and RS-2025-02214715), and the generous supercomputing time provided by the Korea Institute of Science and Technology Information (KISTI). G.H. acknowledges support from the U.S. National Science Foundation under Grant # CBET-2442223. This work used NCSA Delta CPU at University of Illinois Urbana-Champaign through allocation MAT250081 from the Advanced Cyberinfrastructure Coordination Ecosystem: Services & Support (ACCESS) program, which is supported by U.S. National Science Foundation grants #2138259, #2138286, #2138307, #2137603, and #2138296.

# Supplementary Information of "Symbolic Predicate-Guided Language Agents for Inverse Design of Perovskite Oxides"

*Dong Hyeon Mok*[1], *Seoin Back*[2,3,4,*], *Victor Fung*[5,*] *and Guoxiang Hu*[6,7,*]

[1]Department of Chemical and Biomolecular Engineering, Institute of Emergent Materials, Sogang University, Seoul 04107, Republic of Korea

[2]KU-KIST Graduate School of Converging Science and Technology, Korea University, Seoul 02841, Republic of Korea

[3]Department of Integrative Energy Engineering, Korea University, Seoul 02841, Republic of Korea

[4]Institute for Multiscale Matter and Systems (IMMS), Ewha Womans University, Seoul 03760, Republic of Korea

[5] School of Computational Science and Engineering, Georgia Institute of Technology, Atlanta, GA 30332, USA.

[6] School of Materials Science and Engineering, Georgia Institute of Technology, Atlanta, GA 30332, USA.

[7] School of Chemistry and Biochemistry, Georgia Institute of Technology, Atlanta, GA 30332, USA.

AUTHOR INFORMATION

**Corresponding Author**

E-mail: sback@korea.ac.kr, victorfung@gatech.edu, emma.hu@mse.gatech.edu

## Supplementary Note A. Details of Prompts Template

### A.1 System Prompt

The system prompts for the three agents, design, interpretation and curate, were configured according to the target property, the desired optimal value, and the strategy used in each run. Each agent prompt consisted of shared instructions common to all agents and agent-specific instructions tailored to its role. Domain specific language (DSL)-related instructions were included only in the DSL-guided strategy, whereas they were omitted from the prompts used in the other strategies. The detailed prompts are summarized in **Table S1** and **S2**.

**Table S1.** Base system-prompt templates for the three LLM agents. Fields enclosed in curly brackets are dynamically populated according to the target property, target value, strategy, and rule-store statistics used in each run.

| Design Agent System Prompt Base |
|---|
| You are a computational chemist designing oxide perovskite.<br>**{target_property}**<br>**{target_value}**<br><br>Material system: Oxide Perovskite.<br>- Compositions: single ABO3 (Ap=A, Bp=B) or double perovskites (A2BB'O6, AA'B2O6, AA'BB'O6).<br>- X is always O.<br>- Charge neutrality must hold.<br><br>Output schema: the element-based PerovDescription schema.<br>- Fill A, Ap, B, Bp, X. For single perovskites, Ap=A and Bp=B.<br>- Provide a short `name` identifying the composition.<br><br>Rules:<br>- Do NOT duplicate any tested material.<br>- Explore diverse chemistry.<br>- Aim for materials that succeed on as many targets as possible simultaneously.<br><br>For Exploration: novel, diverse compositions.<br>For Exploitation: small variations on top performers (per-target or overall).<br>**{stats_block}** |
| Interpretation Agent System Prompt Base |
| You are a materials data analyst for an oxide perovskite discovery loop.<br>**{target_property}**<br><br>You will be given:<br>- Current iteration materials (formulas + per-target values + per-target success/failure).<br>- Past iteration materials (same).<br>- Current curated rule store (with per-effect precision/lift when verified).<br><br>Your job:<br>1. Produce fresh insights specific to this iteration.<br>2. Propose new or refined `new_rules`. Rules may explicitly discuss multi-target trade-offs.<br><br>Do not judge success/failure yourself, labels are pre-computed per target.<br>**{stats_block}**<br>**{dsl_block}** |
| Curate Agent System Prompt Base |
| You are a rule librarian for a materials discovery loop with possibly multiple targets.<br>You maintain a concise, non-redundant rule store (max **{max_rules}** rules). |

**{targets_property}**

You will be given the current store and newly proposed rules.

Produce the final curated store:
- Merge, drop weak/outdated/redundant, keep well-supported ones.
- Rules may reference multiple targets, preserve such nuance when merging.
- Total must not exceed **{max_rules}**.
**{stats_block}**
**{dsl_block}**

**Table S2.** Dynamically configured prompt components used in the agent system prompts.

<table>
<tr><th colspan="12">{target_property}</th></tr>
<tr><td colspan="3">Binding Energy</td><td colspan="3">Formation Energy</td><td colspan="3">Band Gap</td><td colspan="3">OER Overpotential</td></tr>
<tr><td colspan="3">Binding energies are Gibbs free $\Delta G$ (eV).<br><br>Higher = Weaker binding<br>Lower = Stronger binding</td><td colspan="3">Formation energy is per atom (eV/atom) against Materials Project element references.<br><br>More Negative = More stable</td><td colspan="3">Band gap is in eV (MEGNet -predicted).<br><br>Larger = More insulating<br>Smaller = More metallic<br>0 ≈ metallic / no gap</td><td colspan="3">OER limiting overpotential $\eta_{OER}$ (V) is derived from the 4-step<br><br>OER mechanism (* → *OH → *O → *OOH → $O_2$)<br><br>Lower = Better catalyst<br>0 V is thermodynamic ideal</td></tr>
<tr><th colspan="12">{target_value}</th></tr>
<tr><td colspan="4">Match</td><td colspan="4">Minimize</td><td colspan="4">Maximize</td></tr>
<tr><td colspan="4">Match optimum {value} ± {threshold}</td><td colspan="4">Minimize, success when value < {value}</td><td colspan="4">Maximize, success when value > {value}</td></tr>
<tr><th colspan="12">{stats_block}</th></tr>
<tr><td colspan="12">Some rules show one ‘verified’ line per affected target with stats from accumulated data.<br><br>- target / claimed direction (↑ or ↓): the claim under test for that target.<br>- matched/total: # materials where the predicate is True / total observed.<br>- precision: among matched materials, fraction whose target value is on the claimed direction side of the verifier's split value.<br>- baseline: same fraction over the whole dataset.\n"<br>- lift = precision − baseline. Positive ⇒ predicate enriches for the claimed direction. Negative ⇒ predicate is anti-correlated with the claim.<br>- CI95 = [ci_low, ci_high]: bootstrap 95% CI on lift. If 0 falls inside, the enrichment is not statistically distinguishable from chance.<br>- consistent / mismatch: whether lift > 0 (data supports the claim).<br><br>Split value (per target):<br>- If the target's spec lists `criteria=X`, the split is fixed at X. Precision lift then measure enrichment toward the GOAL-aligned side of X, typically X is set to the success cutoff, so lift > 0 means 'predicate enriches for actually-successful materials' rather than 'below-median materials’.<br>- Otherwise the split is the dataset median (baseline ≈ 0.5 by construction). Lift > 0 means the matched group is enriched below/above the median for that target, which is goal-aligned only when the median itself sits on the right side of the goal. Be cautious when the goal threshold is far from the median.<br><br>Rules whose tier is L2 carry no stats — either no predicate was provided, the predicate failed to parse, or insufficient data was available. Judge them on chemistry intuition.<br><br>Mode-aware interpretation (when split = median, i.e. criteria not set):<br>- For a minimize target: claim='decrease' with positive lift ⇒ predicate enriches the low side, which is the goal direction.<br>- For a maximize target: claim='increase' with positive lift ⇒ predicate enriches the high side, which is the goal direction.</td></tr>
</table>

- For a match target: enrichment direction is goal-aligned only when the target value sits on that side of the dataset median. Combine the stat with chemistry context. A rule with positive lift may not always be moving compositions toward the target.
When `criteria` is set, the goal-alignment is built into the split itself, so lift > 0 directly means 'predicate enriches for the goal side of criteria'.

**{stats_block_tail}**

## {stats_block_tail}

| *Design* | *Interpretation* | *Curate* |
|---|---|---|
| Use these stats to inform candidate selection. Guidelines:<br><br>- Exploitation: prefer compositions that satisfy verified predicates with strong positive lift in the direction you want.<br>- Exploration: prefer compositions that fall outside the matched subgroups of strong predicates, chemical space coverage.<br>- Avoid compositions implied by direction-mismatched predicates.<br>- L2 rules carry no stats. Treat them as qualitative chemistry guidance. | When proposing new_rules:<br>- Avoid restating rules that are already L1-verified with strong consistent lift.<br>- If an existing rule shows direction 'mismatch, propose a refined version. (e.g., add another conjunct to narrow the matched subgroup, or invert the claim).<br>- Wide CI rules just need more data. Do not duplicate them.<br>- Compound predicates (`and`/`or`) often yield sharper enrichment than single feature ones. Prefer compound when chemistry warrants. | Use your judgment when curating:<br>- lift > ~0.2 with CI excluding 0 ⇒ strong evidence; preserve the rule.<br>- Wide CI or lift near 0 ⇒ inconclusive, may need more data, do not drop yet.<br>- Direction mismatch (lift ≤ 0) with CI cleanly negative ⇒ data contradicts the claim, either flip the claim or refine the predicate.<br>- A predicate matching too few materials (small n_matched) ⇒ refine to be less restrictive, or note that more data is needed.<br>- Note your stats-driven decisions in `change_summary` for an audit trail. |

## {dsl_block}

Each rule additionally carries two structured fields in its output schema. When predicate_dsl and at least one entry in `effects` are present, the system automatically validates the rule against accumulated data and reports per-target precision / lift back to you in subsequent iterations.

- predicate_dsl: a DSL predicate (Python-like syntax) over `material` that evaluates to True for the materials this rule applies to. The verifier coerces the result via Python `bool(...)`, so any expression that resolves to truthy/falsy works (bool primitives, comparisons, compound conditions).
- effects: a list of per-target claims. Each entry: target_name (one of [{target_names}]) + claim ('increase' or 'decrease'). You can list multiple targets in one rule's effects — each gets its own verified stats.

Fill these when the rule's applicability is naturally a single bool predicate and it makes directional claims on one or more targets. If the rule mentions concepts outside the DSL (d/p hybridization, Jahn-Teller activity, electronic structure, magnetism, etc.) or is a strategic / operational guideline, leave predicate_dsl null and effects empty. The rule is still kept in the store and shown to the design agent as a qualitative guideline.

When merging rules in a curator role, preserve the (predicate_dsl, effects) pair if either source had them; rewrite predicate_dsl only if the merged rule's applicability genuinely changes.

Recommended building blocks for predicates:\
- has_element(X, site) → bool
- contains_block_metal(site, block) → bool, block in {"3d","4d","5d","4f","5f"}
- numeric_primitive < / > / ≤ / ≥ / == threshold → bool
- combine via `and` / `or` / `not`

Compound predicates (mixing multiple conditions) often yield much sharper enrichment than single-feature predicates. Prefer them when chemistry warrants.

**Supplementary Note B. Details of Symbolic Predicates**

**B.1 Mechanism and Role of Symbolic Predicates**

A design rule suggested by the interpretation agent captures chemical relationships between materials and target properties, but it relies on the reasoning capabilities of the agent and its validity cannot be verified during the loop. To ground such rules in designed data, the interpretation agent is instructed to generate an executable symbolic predicate. Rules involving concepts outside the DSL may remain as qualitative $L_2$ rules without predicates. It commits to a 'Symbolic Predicate' that returns True or False for the materials to which the rule is claimed to apply, covering the space of material symbolic descriptors handled into the loop. Thus, each rule therefore carries their structured fields jointly: 1) a natural language rule statement, 2) a symbolic predicate encoded in chemistry domain specific language (DSL), and 3) one or more target-specific effects specifying whether the proposed predicate is expected to increase or decrease each target property.

The predicate enables per-iteration data-driven validation. For every rule, the framework extracts from the accumulated dataset the subset of materials satisfying the corresponding predicate and evaluate whether their property distribution is consistent with the claimed effect. This procedure provides statistical evidence supporting or contradicting the natural language rule. For example, if a rule states that the presence of Fe increases a target property relative to the overall dataset median, the framework identifies all accumulated materials containing Fe and computes the fraction of those materials whose target property values exceed the dataset median. This fraction is then used to assess whether the DSL supports the proposed rule.

**B.2 List of Symbolic Predicates**

The DSL primitives form an immutable registry presented to the interpretation agent in the system prompt. The registry for DSL partitions into four categories as following tables:

**Table S3.** Categorical primitives available in the chemistry-specific DSL. Each primitive returns either a Boolean value indicating whether a compositional condition is satisfied or a list of elements occupying the specified crystallographic sublattice.

| Predicate | Return | Semantics |
|---|---|---|
| has_element(*e*, site) | bool | True if *e* occupies the requested site |
| contains_block_metal(site, block) | bool | True if some element on site belongs to the given orbital block |
| A_site_elements(materials) | list | Distinct symbols at the A site |
| B_site_elements(materials) | list | Distinct symbols at the B site |

**Table S4.** Numerical primitives available in the chemistry-specific DSL. The primitives return elemental fractions, composition-weighted property means, maximum or minimum elemental-

property values, or the Goldschmidt tolerance factor. Numerical outputs can be combined with comparison operators to construct Boolean symbolic predicates.

| Predicate | Return |
|---|---|
| fraction(e, site) | float in [0,1] |
| mean_property(site) | composition-weighted mean |
| max_property(site) | maximum over site elements |
| min_property(site) | minimum over site elements |
| tolerance_factor(material) | Goldschmidt factor |

**Table S5.** Elemental descriptors exposed to the numerical DSL primitives. A total of 46 elemental properties is exposed via the property argument of mean, max and min_property predicates.

| Category | Members |
|---|---|
| Identity (7) | atomic_weight, atomic_number, period, group, quantum_number, Mendeleev_numer, family |
| Orbital counts (10) | valence_{s, p, d, f}_electron_count, unfilled_{s, p, d, f}_electron_count, valence_electron_count, outer_shell_electron_count |
| Energy & electronegativity (8) | metallic_valences, $Z_{eff}$, first_bohr_radius, ionization_energy, electron_affinity, allred_rockow_EN, Nagle,EN, Ghosh_EN |
| Radii (11) | atomic_radius, covalent_radius, ionic_radius, effective_ionic_radius, miracle_radius, van_der_waals_radius, zunger_raddi_sum, crystal_radius, covalent_CSD_radius, slater_radius, orbital_radius |
| Thermal (8) | melting_point, boiling_point, density, specific_heat, heat_of_fusion, heat_of_vaporization, heat_of_atomization, thermal_conductivity |
| Misc (2) | polarizability, oxidation_state |

**Table S6.** Logical, comparison, arithmetic, and mathematical operators supported by the chemistry-specific DSL. The available operators allow the interpretation agent to combine categorical and numerical primitives into compound Boolean predicates and to express arithmetic relationships among elemental descriptors.

| Operator class | Members |
|---|---|
| Boolean | and, or, not |
| Comparison | $<, \leq, >, \geq, ==, \neq$ |
| Arithmetic | +, −, ×, ÷, **, % |
| Math helpers | abs, min, max |

### B.3 Rule Evaluation Metrics

For each stored at a given iteration, the framework evaluates each claimed effect on the target properties by computing partition-based statistics. These statistics summarize whether the accumulated data supports the claimed rule that materials satisfying the corresponding DSL exhibit the expected increase or decrease in a target property.

#### B.3.1 Partition and Claim Indicator

At each iteration, the accumulated dataset is restricted to materials for which the relevant target property has been numerically evaluated. The symbolic predicate associated with a rule is then applied to each material and returns a Boolean outcome indicating whether the material satisfies the predicate (i.e., falls within the predicate's applicability domain). In parallel, and independently of the predicate outcome, each material is assigned a claim-consistency outcome based on whether its target value lies on the expected side of a target-specific split value. For an increase claim, a material is considered consistent with the claim if its target value is higher than the split value. For a decrease claim, a material is considered consistent with the claim if its target value is lower than the split value.

Using the predicate outcome, the dataset is divided into two partitions. The matched partition contains materials that satisfy the symbolic predicate, whereas the unmatched partition contains materials that do not satisfy it. This partition does not yet incorporate the claim direction, and the interaction between predicate and claim is captured separately by the following metrics.

#### B.3.2 Precision, Baseline, and Lift

For each rule, target property and claimed direction, the verifier reports three quantities, precision, baseline and lift. Precision is defined as the fraction of matched materials whose target values are consistent with the claimed direction. Baseline is the corresponding success rate over the full evaluated dataset, representing the rate expected from uniformly sampling materials without applying the predicate. Lift is the difference between the matched precision and the baseline.

A positive lift indicates that materials satisfying the predicate are enriched in the claimed direction beyond the dataset-based expectation. In contrast, a negative lift indicates that the predicate is anti-correlated with the claimed effect.

#### B.3.3 Bootstrap Confidence Interval for Lift

To estimate the statistical uncertainty associated with lift, the framework reports a 95% bootstrap confidence interval. The verifier repeatedly resamples the evaluated materials with replacement and recomputes precision, baseline and lift for each resampled dataset. The reported confidence interval is obtained from the lower and upper percentiles of the bootstrap lift distribution. To ensure reproducibility, the random seed for bootstrapping is derived deterministically from a hash of the symbolic predicate, target property and claimed effect direction. As a result, applying the same predicate to the same data produces identical confidence intervals across repeated iteration.

### B.3.4 Direction Consistency and Tier Assignment

Each verified statistics record includes a direction consistency flag. This flag is set to true when the empirical lift supports the direction claimed by the rule. A false value indicates that the data contradicts the claimed effect. Such contradictions guide the curate agent in subsequent iterations, where the agent may reverse the claimed direction, narrow the predicate to exclude counterpart, or remove the rule entirely.

Rules are assigned to one of two tiers according to whether their effects can be statistically evaluated. $L_1$ rules are statistically evaluated rules. A rule is assigned to $L_1$ when its symbolic predicate can be successfully parsed and at least one of its claimed effects produces computable statistics under the inclusion thresholds. Importantly, $L_1$ assignment does not by itself imply that the rule is chemically correct or strongly supported; rather, it indicates that the rule has been evaluated and is accompanied by its statistical record. The strength of support should be interpreted from the matched-sample count, precision, baseline, lift, confidence interval, and direction consistency flag.

$L_2$ rules are unevaluated or unverified rules. A rule is assigned to $L_2$ when it lacks a predicate, when its predicate fails to parse, when all of its effects are excluded by the inclusion thresholds, or when no effects are declared. These rules are still provided to the design agent, but only as qualitative guidelines without statistical adjudication.

The tier label serves as the summary signal of predicate trustworthiness available to the design agent. The design agent may then weigh $L_1$ and $L_2$ rules differently during its own reasoning process.

### B.3.5 Multi-Target Verification

A distinctive feature of the framework is that each symbolic predicate can support multiple target-specific claimed rules. The predicate is evaluated only once for each and the resulting matched and unmatched partitions are reused across all declared effects. Each effect is then verified independently with its own target property, claimed direction, and statistical record.

For example, a single rule may claim that B-site transition metals lower the oxygen evolution overpotential and increase the band gap. The framework can assign separate verified statistics records to these two claims. The curate agent can then revise the claims selectively, such as removing the band gap claim while preserving the overpotential claim, without evaluating the symbolic predicate again.

### B.3.6 Store-Level Aggregate Report

At the end of each iteration, the verifier returns an aggregate summary of the rule store. This summary includes the total number of rules, the number of $L_1$ and $L_2$ rules, the number of rules that submitted a symbolic predicate, the number of predicates that failed to parse and were automatically cleared, the cumulative number of per-material predicate evaluation errors, and the number of the rule effect claims with non-positive lift.

This aggregate report is logged and included in the design agent prompt as a coarse-grained quality signal for the curated rule store. It complements the detailed per-rule statistics that are provided alongside individual rules.

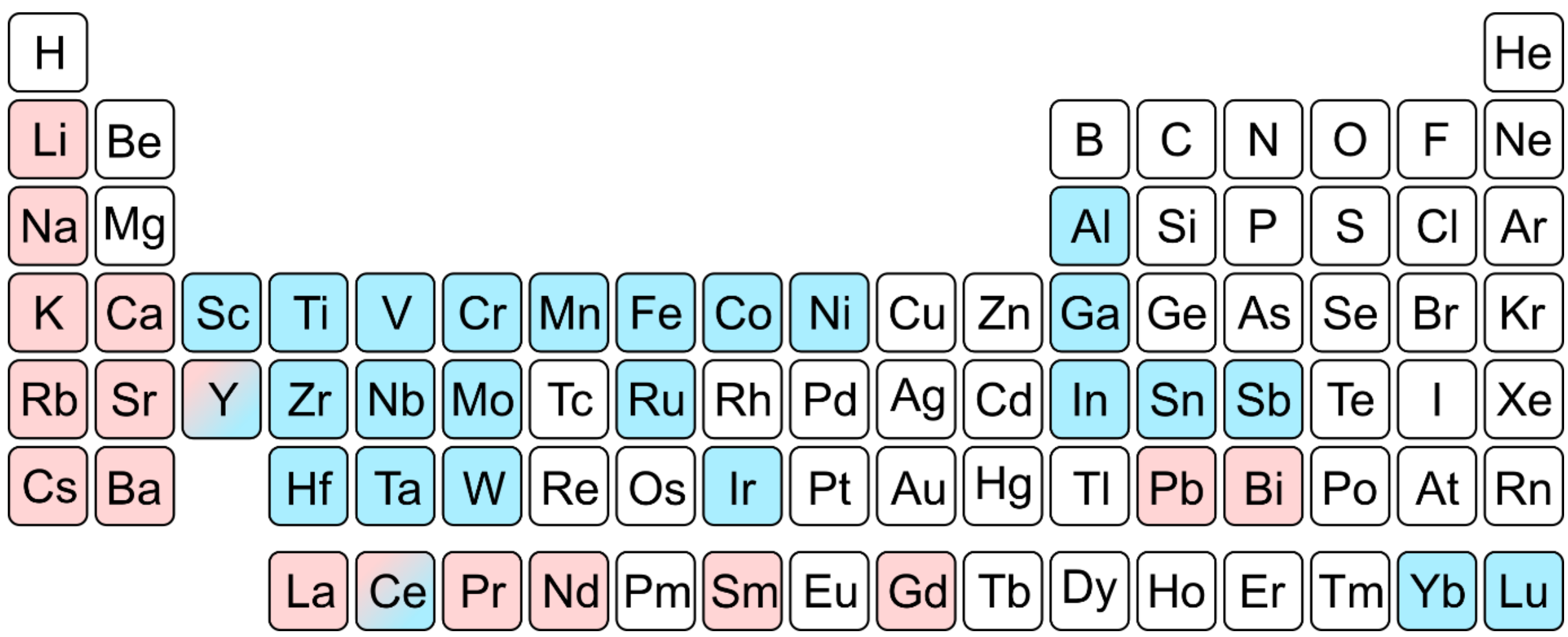

**Figure S1.** Elemental search space used for oxide perovskite design. Periodic table representation of the elements allowed at the A/A′ and B/B′ sites. Red and blue backgrounds indicate elements available for the A-site and B-site sublattices, respectively, and elements marked with both colors are permitted at either site.

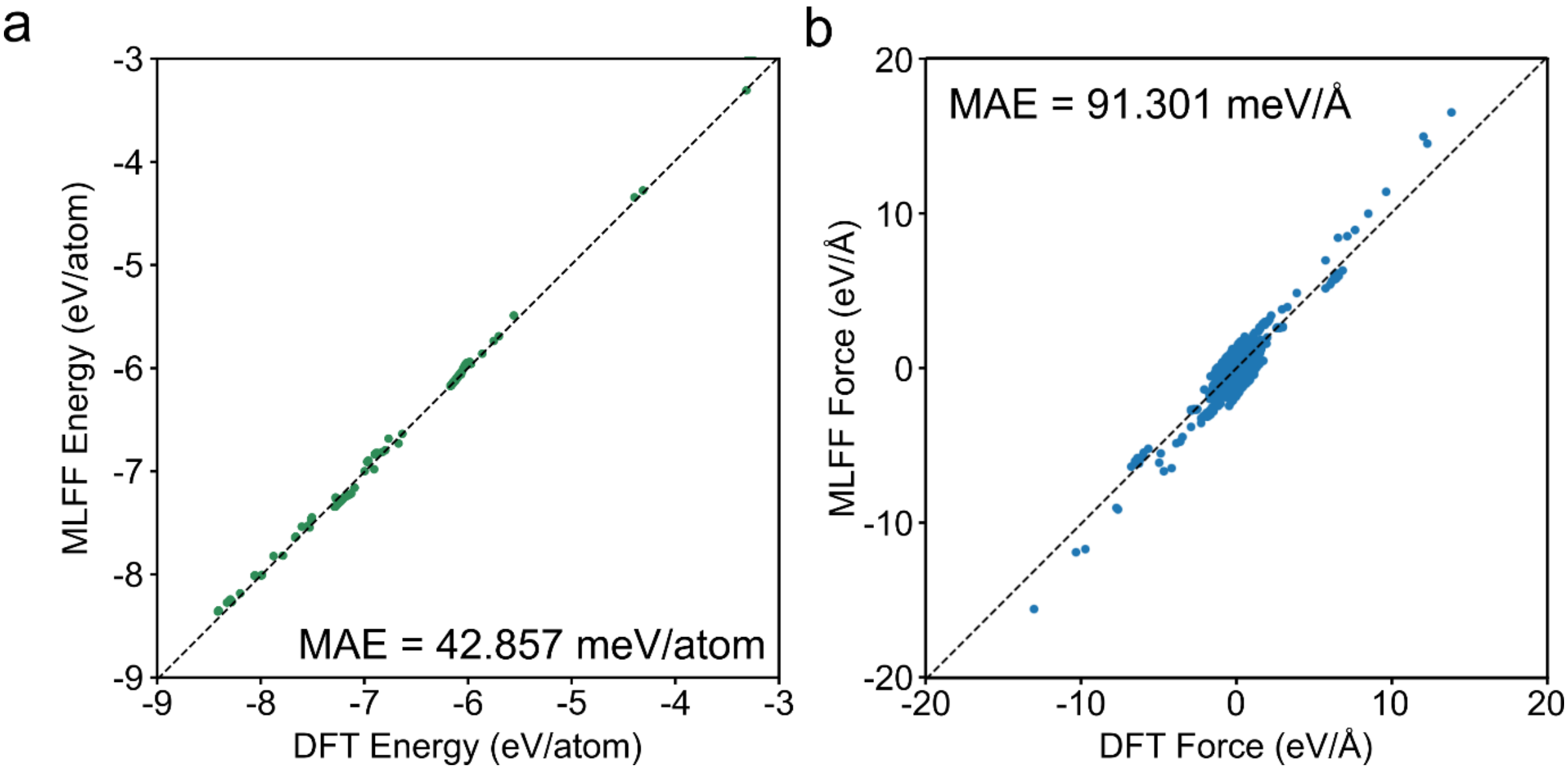

**Figure S2.** Parity plots comparing DFT-calculated and UMA-predicted (a) total energies and (b) atomic forces for the bulk validation set. UMA predictions were obtained using the OMat task.

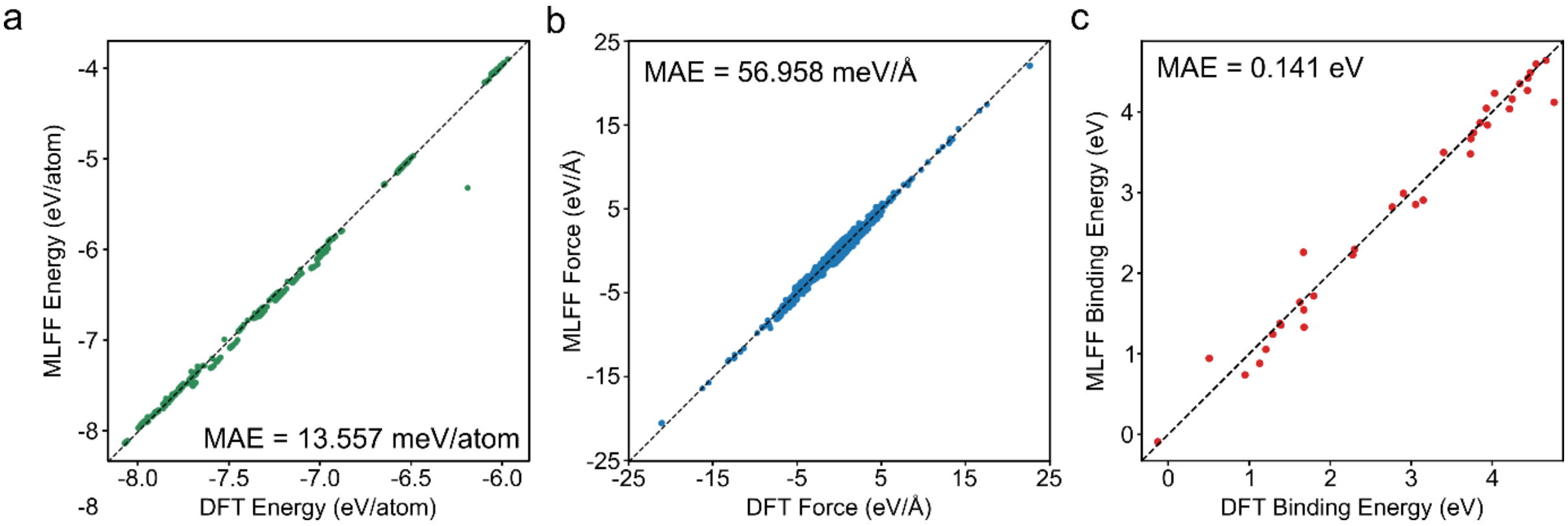

**Figure S3.** Parity plots comparing DFT-calculated and UMA-predicted (a) total energies, (b) atomic forces, and (c) binding Gibbs free energies of OER intermediates for the surface validation set excluding La-containing perovskites.

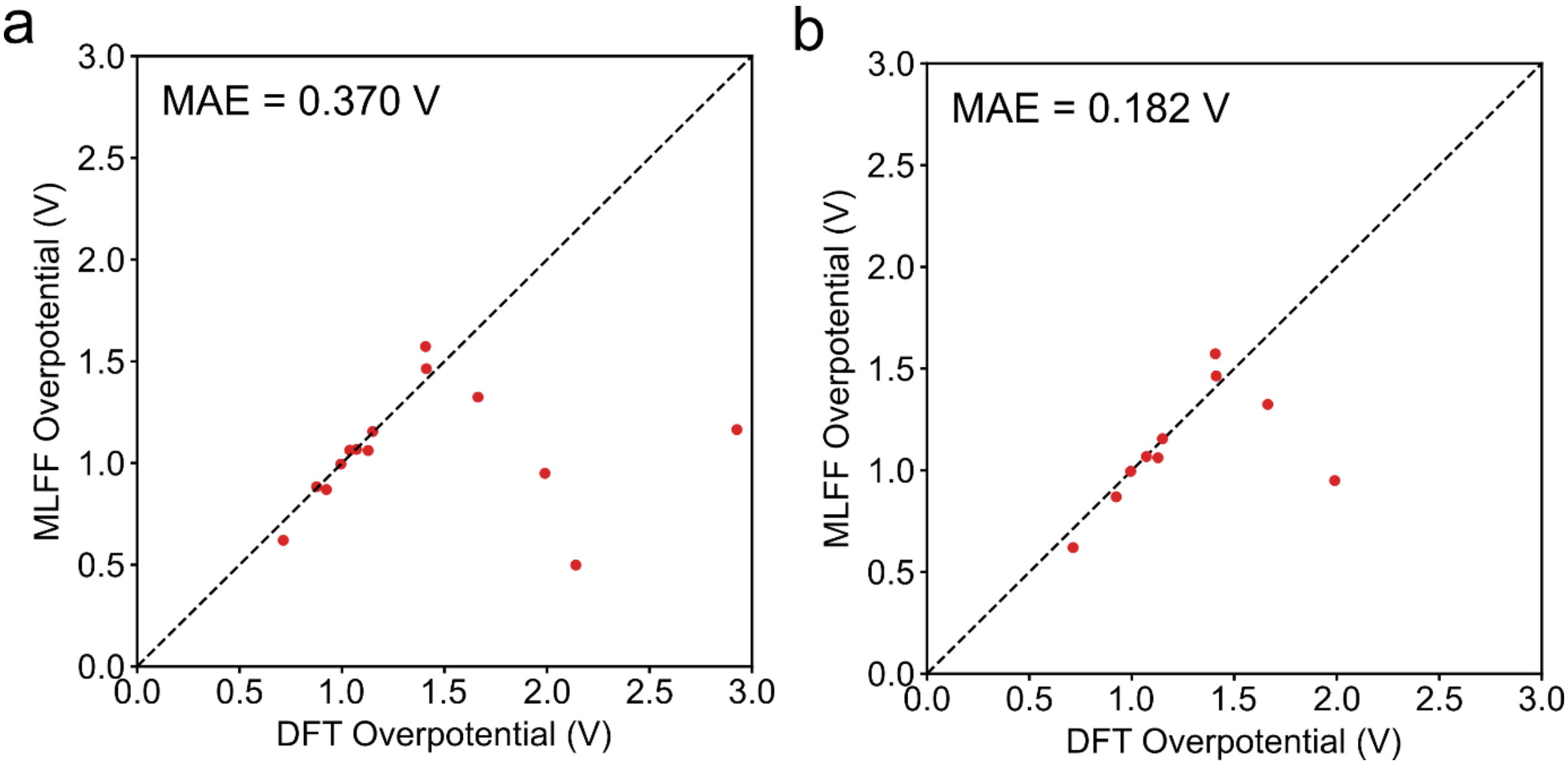

**Figure S4.** Parity plots comparing DFT-calculated and MLFF-predicted overpotential of double perovskite surface structures (a) with La-containing perovskites and (b) without La-containing perovskites using UMA.

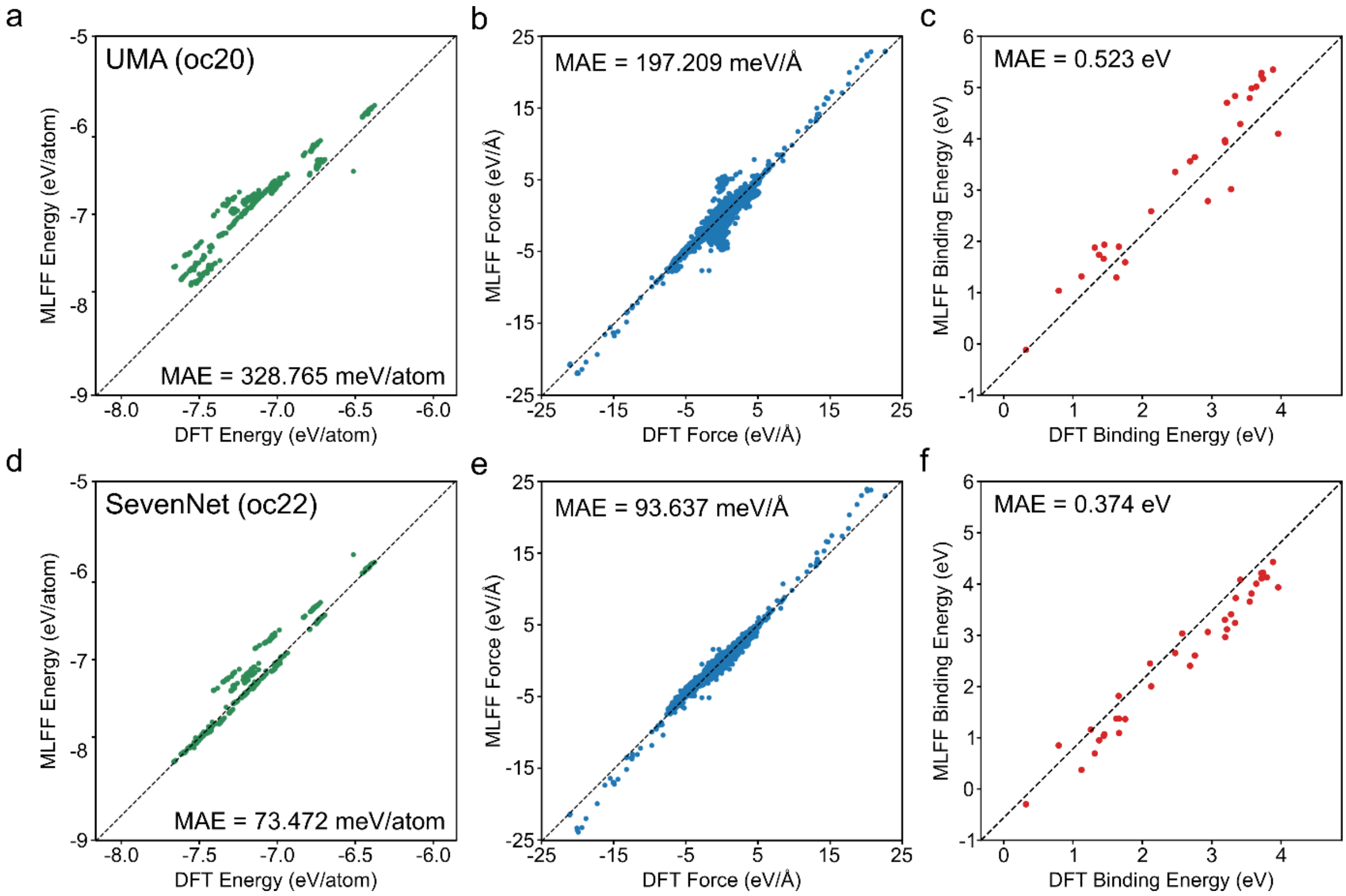

**Figure S5.** Parity plots comparing DFT results with predictions from (a–c) UMA using the OC20 task and (d–f) SevenNet-omni using the OC22 task. The evaluated quantities are total energy, atomic force, and Gibbs free energy, respectively.

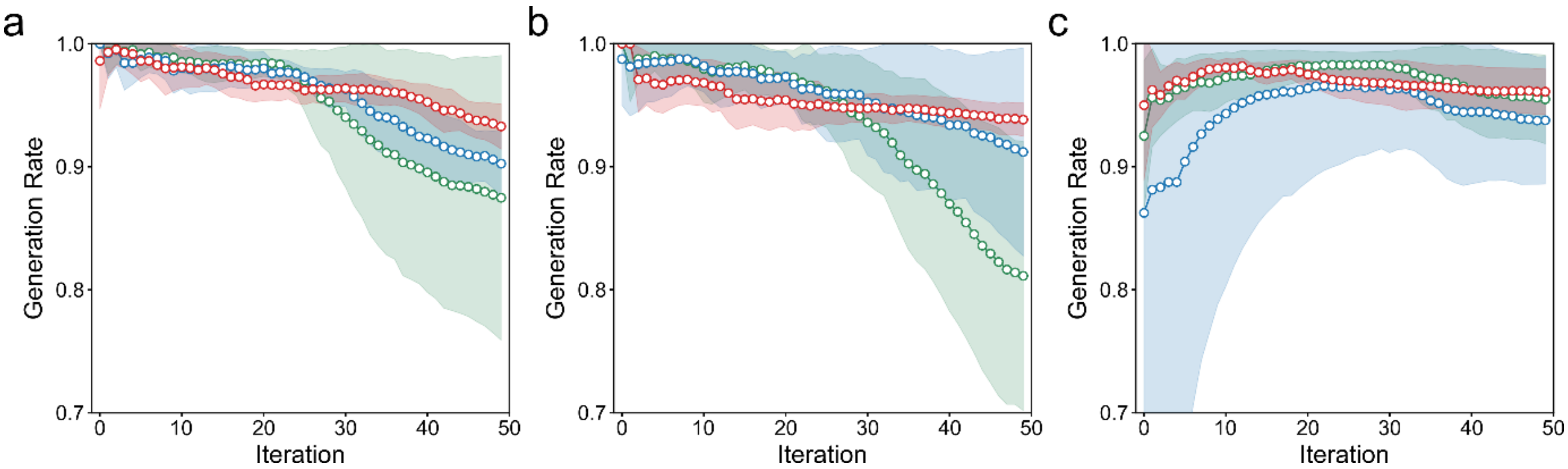

**Figure S6.** Generation rate during the 50 iteration design runs. The generation rate was defined as the fraction of non-duplicate materials successfully generated for exploitation relative to the number of exploitation candidates requested at each iteration. Results are shown for (a) the joint band gap/formation energy target, (b) the OER overpotential target, and (c) the $\Delta G_{*OH}$ target.

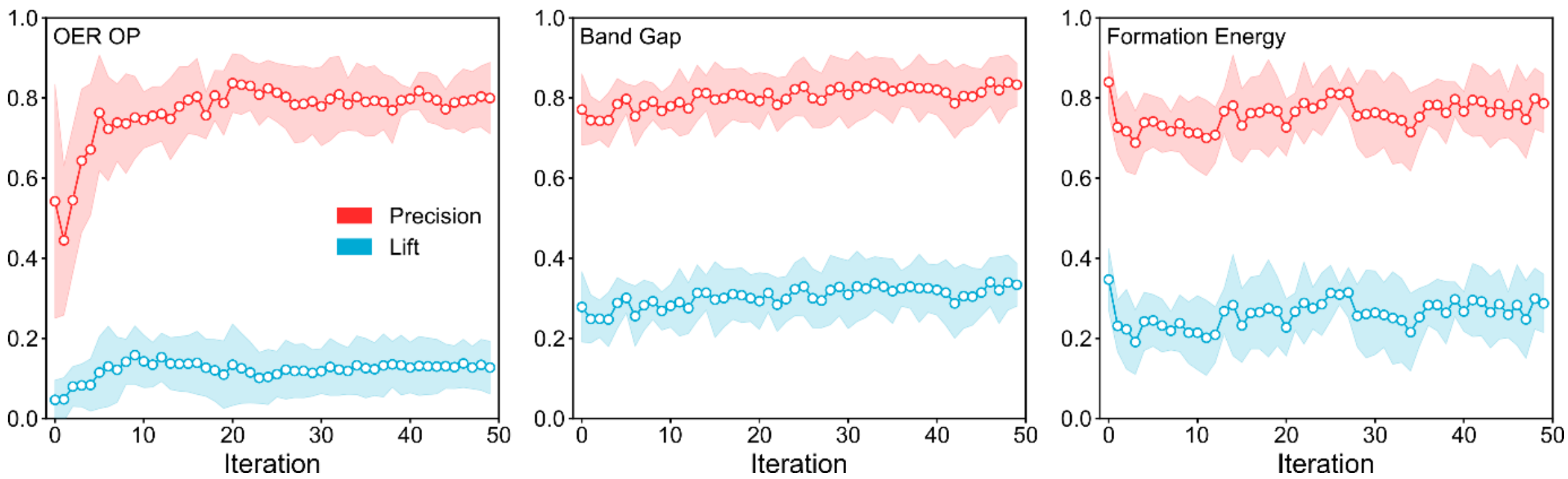

**Figure S7.** Evolution of symbolic predicate statistics for the two primary design tasks. Mean precision and lift are shown as functions of iteration for the OER-overpotential and band gap/formation energy objectives. Values were calculated from the predicates accumulated in the rule store and averaged over 10 independent runs. Shaded regions indicate the standard deviations across runs.

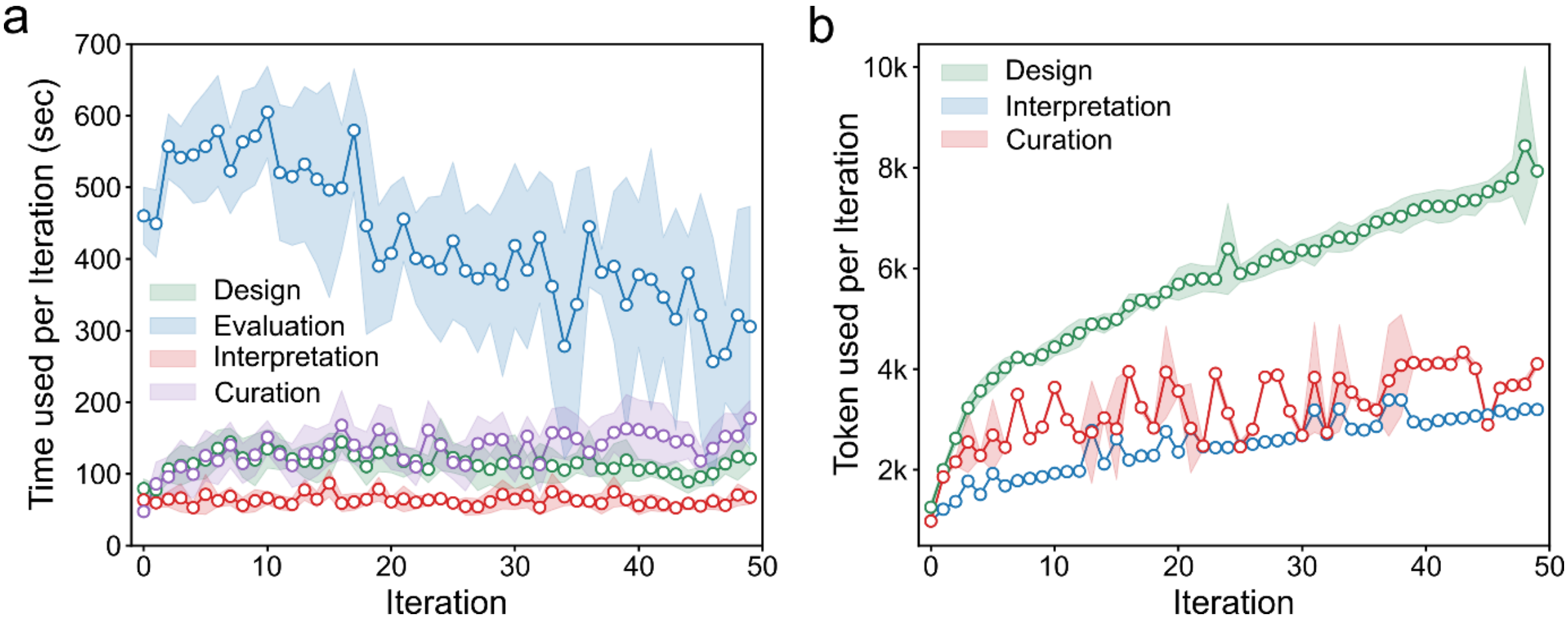

**Figure S8.** Computational and language model costs of the DSL-guided design loop. (a) Mean wall clock time spent per iteration on material design, property evaluation, data interpretation, and rule curation. Property evaluation includes MLFF-based structural optimization and property calculation, whereas the other stages primarily consist of agent reasoning and rule processing. (b) Total number of input and output tokens consumed per iteration by the design, interpretation, and curate agents. Curves are averaged over 10 independent runs, and shaded regions indicate the standard deviations across runs.

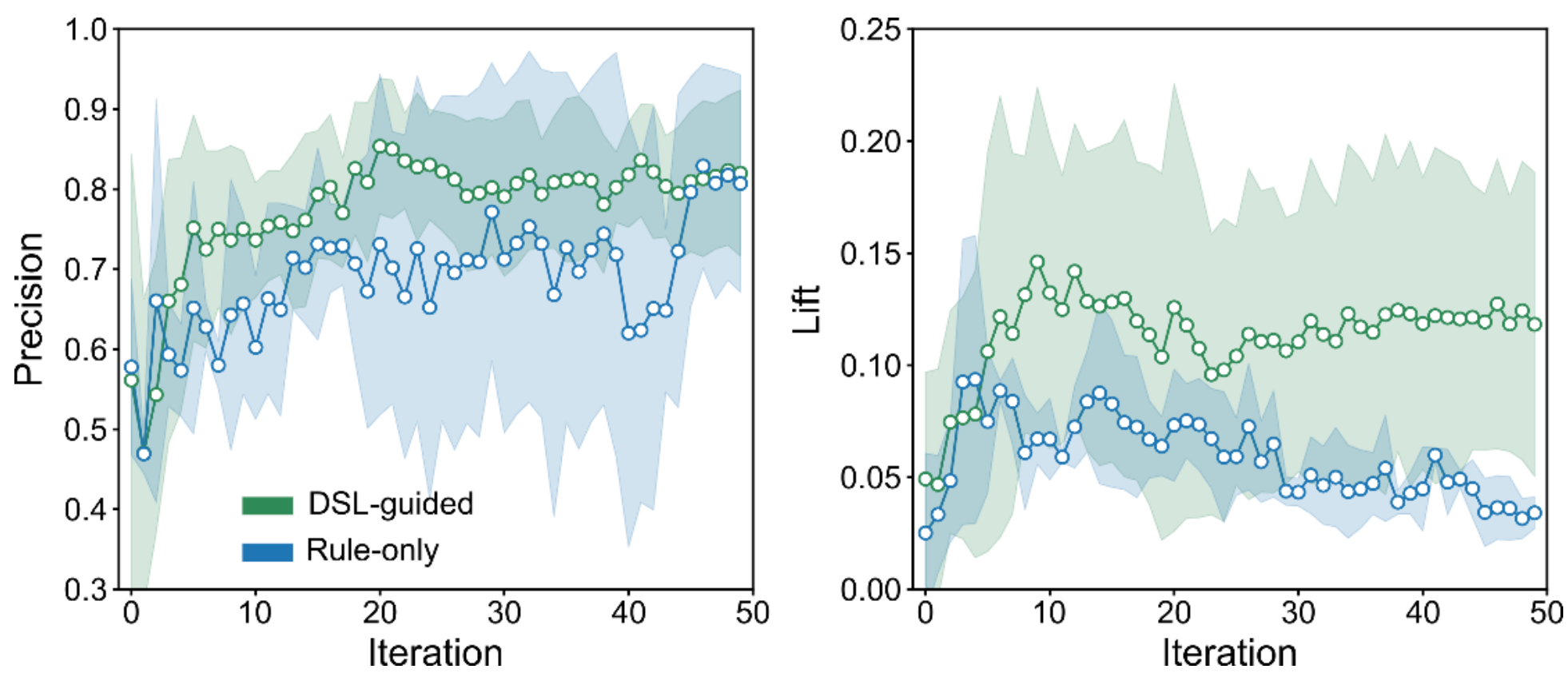

**Figure S9.** Evolution of post hoc predicate statistics for the DSL-guided and rule-only strategies under the OER overpotential objective. (a) Precision and (b) lift of the symbolic predicates as functions of design iteration. For the rule-only strategy, the natural language rules were converted into predicates solely for post hoc analysis using the same formalization procedure, and the resulting statistics were not provided to any agent during the design loop. Curves are averaged over 10 independent runs, and shaded regions indicate the standard deviations across runs.

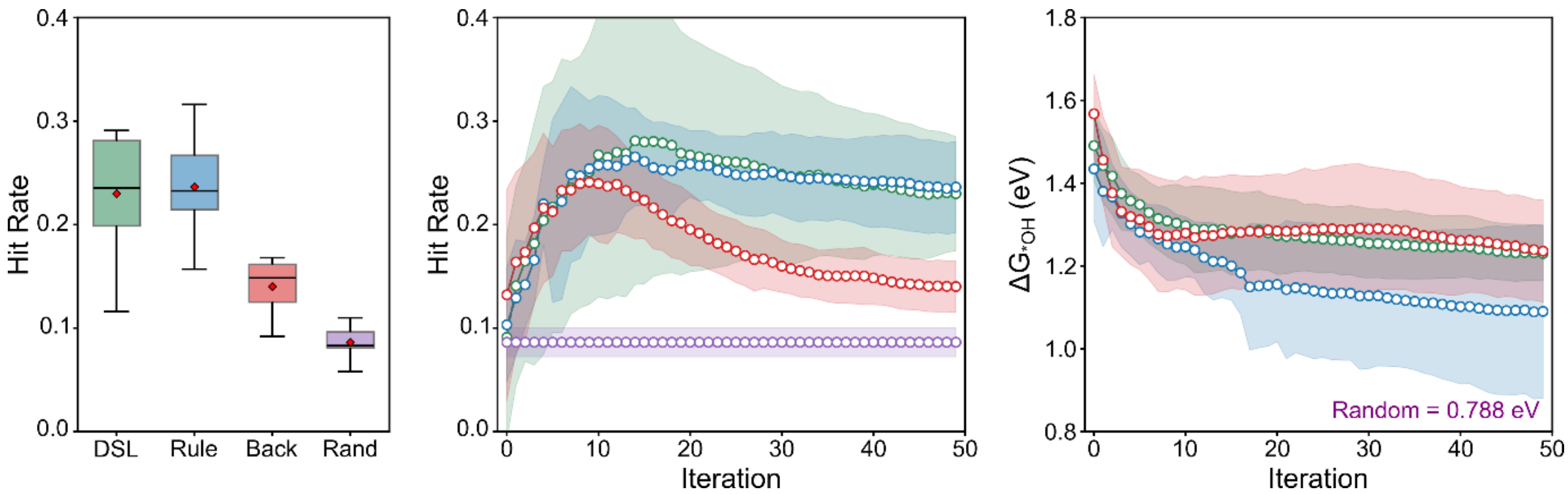

**Figure S10**. Inverse-design performance for the $\Delta G_{*OH}$ objective**.** The target range was defined as 1.13–1.33 eV around the ideal value of 1.23 eV. The left panel shows the distributions of the final iteration hit rates across 10 independent runs. The middle panel shows the mean hit rate as a function of iteration, and the right panel shows the corresponding mean $\Delta G_{*OH}$. Shaded regions indicate the standard deviations across runs.

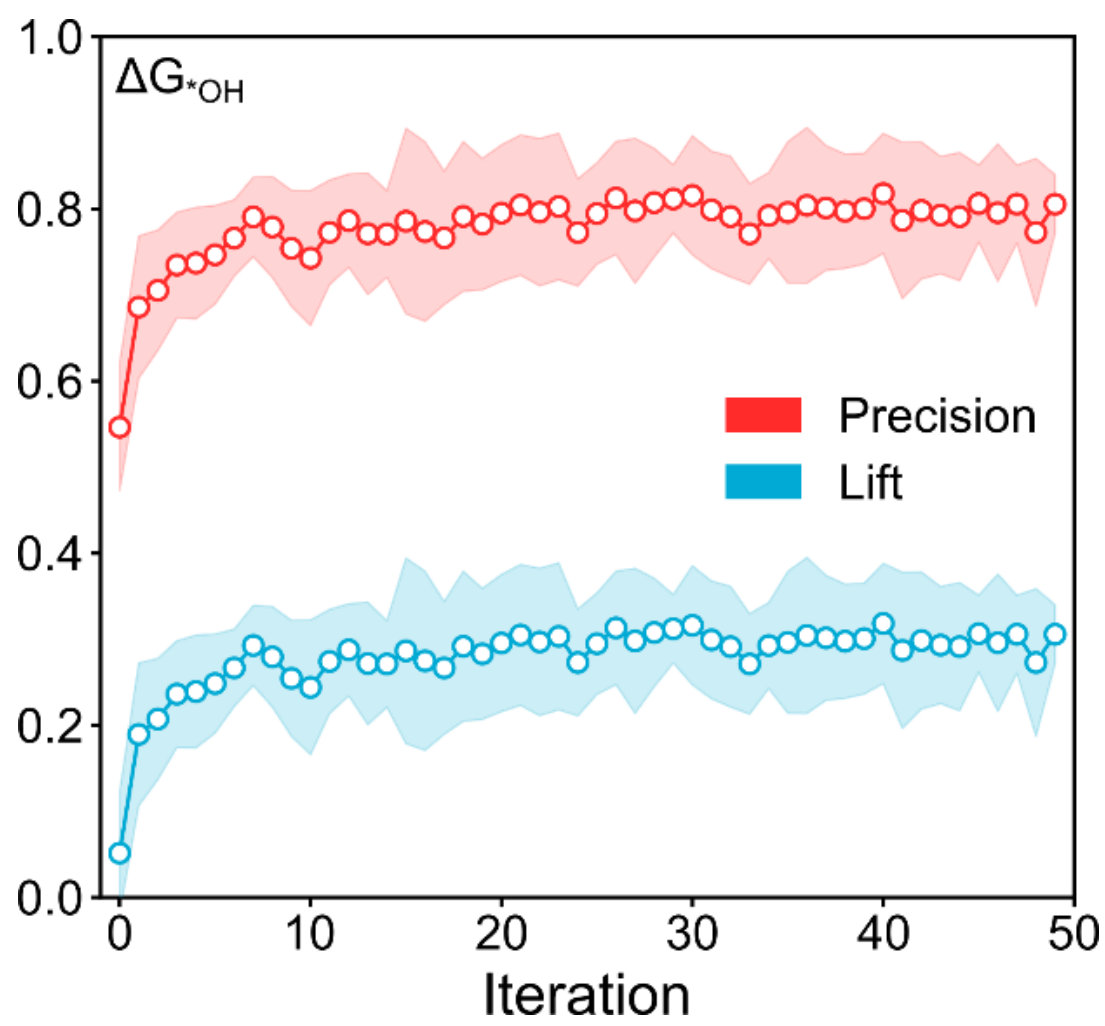

**Figure S11.** Evolution of symbolic predicate statistics under the $\Delta G_{*OH}$ objective. Mean precision and lift are shown as functions of design iteration for predicates accumulated in the DSL-guided rule store. Values were averaged over 10 independent runs, and shaded regions indicate the standard deviations across runs.

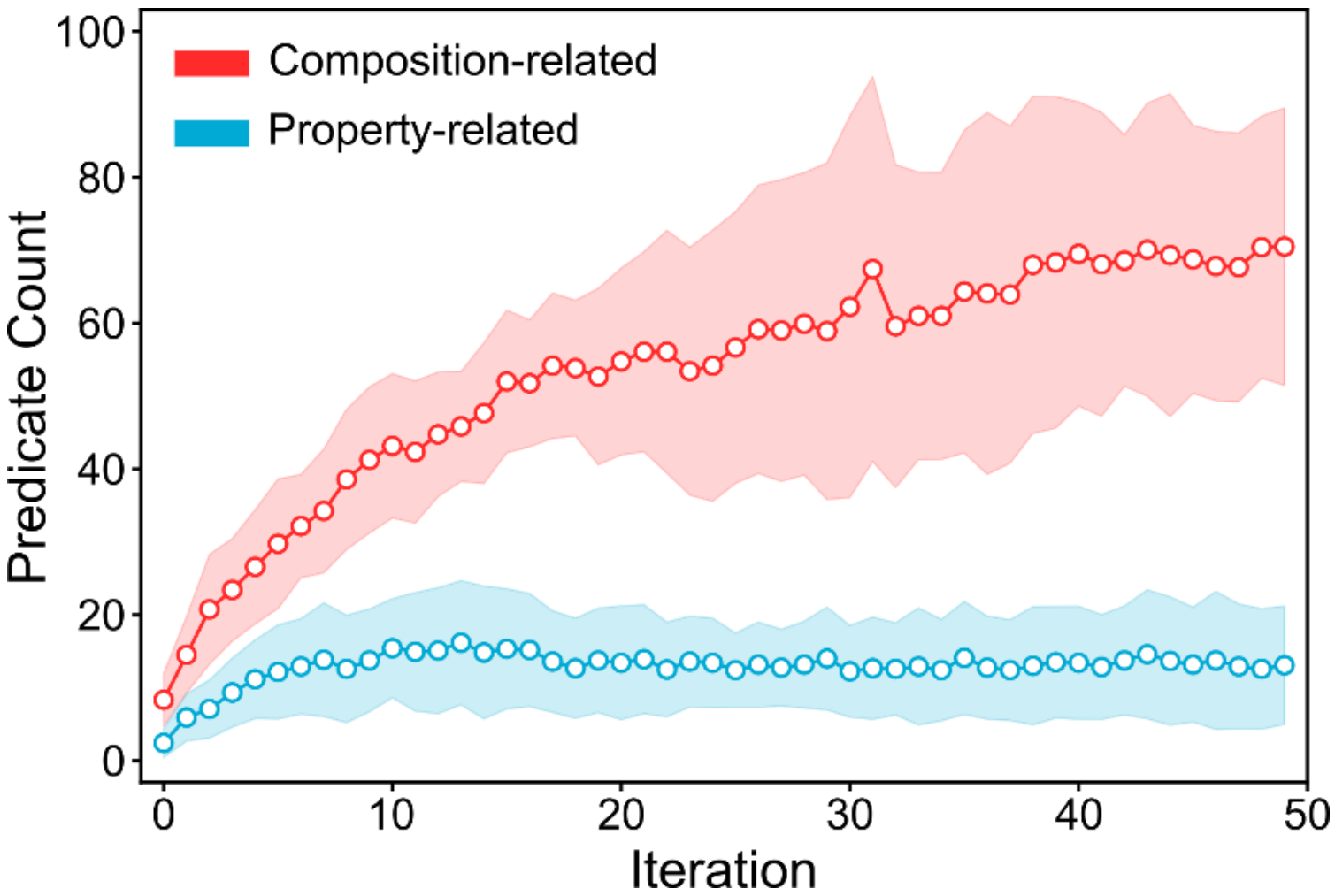

**Figure S12.** Evolution of the representational content of the symbolic predicates under the OER overpotential objective. The numbers of composition-related predicates, which explicitly reference elemental identity or site fraction, and property-related predicates, which use elemental descriptors, are shown as functions of design iteration. Values were averaged over 10 independent runs, and shaded regions indicate the standard deviations across runs.

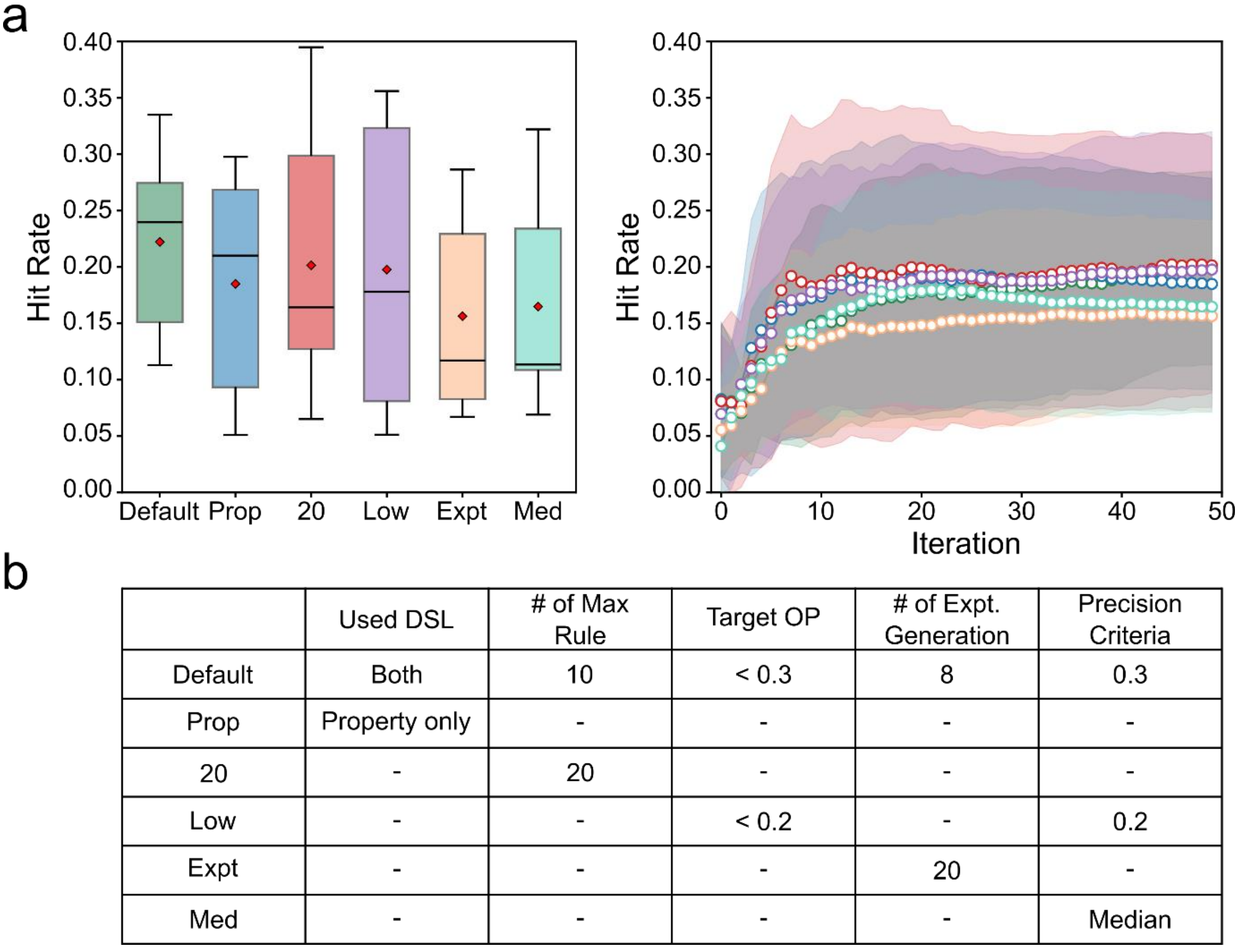

| | Used DSL | # of Max Rule | Target OP | # of Expt. Generation | Precision Criteria |
|---|---|---|---|---|---|
| Default | Both | 10 | < 0.3 | 8 | 0.3 |
| Prop | Property only | - | - | - | - |
| 20 | - | 20 | - | - | - |
| Low | - | - | < 0.2 | - | 0.2 |
| Expt | - | - | - | 20 | - |
| Med | - | - | - | - | Median |

**Figure S13.** Effects of framework and DSL hyperparameters on OER inverse design performance**.** (a) Distribution of the final iteration hit rate and mean hit rate for the default setting and five modified configurations. Shaded regions indicate the standard deviations across 10 independent runs. (b) Hyperparameter settings used in each configuration. "Default" uses the complete DSL, a maximum of 10 stored rules, an OER-overpotential target and statistical criterion of 0.3 V, and eight exploitation candidates per iteration. "Prop" restricts the DSL to elemental property-based predicates. "20" increases the maximum rule store size to 20. "Low" lowers the OER target and criterion to 0.2 V. "Expt" increases the number of exploitation candidates to 20 per iteration, and "Med" replaces the fixed 0.3 V statistical criterion with a dataset median split. Unspecified parameters retain their default values.

**Table S7.** DFT-calculated and UMA-predicted Gibbs free energies for the surface validation set

| | $\Delta G_{*O}$ (eV) | | | $\Delta G_{*OH}$ (eV) | | | $\Delta G_{*OOH}$ (eV) | | |
|---|---|---|---|---|---|---|---|---|---|
| Formula | DFT | MLFF | Error | DFT | MLFF | Error | DFT | MLFF | Error |
| $SrLaMgNbO_6$ | 4.543 | 4.594 | 0.051 | 2.301 | 2.297 | 0.004 | 4.664 | 4.64 | 0.025 |
| $LaAlO_3$ | 3.942 | 3.838 | 0.104 | 1.674 | 1.545 | 0.129 | 4.436 | 4.266 | 0.169 |
| $BaZrO_3$ | 3.928 | 4.045 | 0.117 | 1.290 | 1.242 | 0.048 | 4.340 | 4.350 | 0.010 |
| $KNbO_3$ | 3.149 | 2.905 | 0.244 | 1.206 | 1.055 | 0.151 | 4.274 | 4.159 | 0.087 |
| $BiFeO_3$ | 3.400 | 3.498 | 0.098 | 0.507 | 0.944 | 0.437 | 4.030 | 4.231 | 0.202 |
| $Ca_2FeReO_6$ | 1.670 | 2.260 | 0.589 | -1.412 | 0.531 | 1.943 | 1.549 | 3.722 | 2.172 |
| $CaMnO_3$ | 3.732 | 3.479 | 0.253 | 1.129 | 0.881 | 0.248 | 2.279 | 2.226 | 0.052 |
| $YCrO_3$ | 1.678 | 1.329 | 0.349 | -0.128 | -0.090 | 0.039 | 2.766 | 2.820 | 0.054 |
| $LaFeO_3$ | 4.396 | 3.252 | 1.144 | 3.220 | 1.073 | 2.147 | 4.765 | 4.119 | 0.646 |
| $SrTiO_3$ | 3.771 | 3.741 | 0.030 | 1.392 | 1.356 | 0.035 | 4.442 | 4.419 | 0.023 |
| $BaLaFe_2O_6$ | 3.850 | 3.865 | 0.015 | 1.626 | 1.640 | 0.014 | 2.904 | 2.990 | 0.086 |
| $BaLaFeTiO_6$ | 5.953 | 4.112 | 1.841 | 1.796 | 1.718 | 0.079 | 5.498 | 3.854 | 1.644 |
| $PbTiO_3$ | 3.737 | 3.667 | 0.070 | 1.380 | 1.375 | 0.005 | 4.471 | 4.487 | 0.016 |
| $NaTaO_3$ | 3.055 | 2.852 | 0.204 | 0.950 | 0.738 | 0.212 | 4.213 | 4.037 | 0.176 |

**Table S8.** Run level inverse design performance for the three design tasks. Hit rates are reported for 10 independent runs using the DSL-guided, rule-only, and background strategies. Target criteria for $\Delta G_{*OH}$, Band gap ($E_g$) & Formation energy ($E_f$) and OER overpotential ($\eta_{OER}$) were set to 1.13 eV ≤ $\Delta G_{*OH}$ ≤ 1.33 eV, 2.0 eV ≤ $E_g$ and $E_f$ ≤ -2.0 eV/atom and $\eta_{OER}$ ≤ 0.3 V.

| Target Property | $\Delta G_{*OH}$ | | | $E_g$ & $E_f$ | | | $\eta_{OER}$ | | |
|---|---|---|---|---|---|---|---|---|---|
| Run | DSL | Rule | Back | DSL | Rule | Back | DSL | Rule | Back |
| 1 | 0.255 | 0.316 | 0.168 | 0.688 | 0.657 | 0.480 | 0.236 | 0.155 | 0.180 |
| 2 | 0.174 | 0.241 | 0.163 | 0.705 | 0.518 | 0.384 | 0.123 | 0.114 | 0.122 |
| 3 | 0.116 | 0.270 | 0.129 | 0.680 | 0.698 | 0.451 | 0.211 | 0.202 | 0.113 |
| 4 | 0.289 | 0.157 | 0.153 | 0.659 | 0.641 | 0.489 | 0.072 | 0.147 | 0.133 |
| 5 | 0.207 | 0.224 | 0.165 | 0.664 | 0.647 | 0.526 | 0.292 | 0.098 | 0.127 |
| 6 | 0.285 | 0.257 | 0.092 | **0.786** | 0.694 | 0.479 | 0.174 | 0.062 | 0.176 |
| 7 | 0.196 | 0.188 | 0.124 | 0.660 | 0.694 | 0.481 | 0.281 | 0.197 | 0.178 |
| 8 | 0.269 | 0.219 | 0.144 | 0.668 | 0.644 | 0.551 | **0.302** | 0.147 | 0.080 |
| 9 | **0.291** | 0.279 | 0.157 | 0.584 | 0.595 | 0.603 | 0.223 | 0.160 | 0.114 |
| 10 | 0.216 | 0.213 | 0.106 | 0.720 | 0.637 | 0.486 | 0.216 | 0.200 | 0.132 |
| Average | 0.230 | **0.236** | 0.140 | **0.681** | 0.642 | 0.493 | **0.213** | 0.148 | 0.135 |

**Table S9.** Run level absolute inverse design performance for the three design tasks. The total numbers of unique target-satisfying materials discovered over 50 iterations are reported for 10 independent runs using the DSL-guided, rule-only, and background strategies. Target criteria for $\Delta G_{*OH}$, Band gap ($E_g$) & Formation energy ($E_f$) and OER overpotential ($\eta_{OER}$) were set to 1.13 eV $\leq \Delta G_{*OH} \leq$ 1.33 eV, 2.0 eV $\leq E_g$ and $E_f \leq$ -2.0 eV/atom and $\eta_{OER} \leq$ 0.3 V.

| Target Property | $\Delta G_{*OH}$ | | | $E_g$ & $E_f$ | | | $\eta_{OER}$ | | |
|---|---|---|---|---|---|---|---|---|---|
| Run | DSL | Rule | Back | DSL | Rule | Back | DSL | Rule | Back |
| 1 | 91 | **124** | 65 | 196 | 230 | 178 | 77 | 54 | 68 |
| 2 | 67 | 89 | 65 | 268 | 189 | 138 | 45 | 42 | 46 |
| 3 | 45 | 100 | 49 | 225 | 254 | 169 | 75 | 74 | 43 |
| 4 | 115 | 61 | 59 | 244 | 216 | 180 | 84 | 56 | 49 |
| 5 | 76 | 76 | 62 | 202 | 229 | 202 | 74 | 38 | 49 |
| 6 | 108 | 97 | 35 | **301** | 254 | 182 | 62 | 24 | 66 |
| 7 | 78 | 63 | 47 | 241 | 239 | 179 | **97** | 53 | 65 |
| 8 | 106 | 85 | 57 | 259 | 237 | 43 | 80 | 56 | 30 |
| 9 | 114 | 110 | 59 | 229 | 141 | 222 | 63 | 60 | 43 |
| 10 | 78 | 84 | 41 | 185 | 237 | 185 | 22 | 77 | 49 |
| Average | 87.8 | **88.9** | 53.9 | **238.0** | 222.6 | 167.8 | **67.9** | 53.4 | 50.8 |

**Table S10.** Representative natural language design rules and corresponding symbolic predicates obtained under the OER overpotential objective. The examples illustrate how Co-containing B-site compositions, Ru/Ir-containing B-site environments, A-site elemental motifs, and related compositional constraints were encoded as executable predicates for statistical evaluation during the design loop.

| Natural Language Rule | Symbolic Predicate |
| --- | --- |
| Co on B and (Ru or Ir) on B, B-site mean d_filling in [5.0,6.2], limited A-site alkali (<0.30 total Li/Na/K/Rb/Cs), explicit A-site positive-charge buffering (Ca or moderate Pb/Bi or moderate lanthanide fraction), and no Ni on B -> lower OER_OP | has_element("Co", site="B") and (has_element("Ru", site="B") or has_element("Ir", site="B")) and mean_property("d_filling", "B") >= 5.0 and mean_property("d_filling", "B") <= 6.2 and (fraction("Li", site="A") + fraction("Na", site="A") + fraction("K", site="A") + fraction("Rb", site="A") + fraction("Cs", site="A")) < 0.30 and ((has_element("Ca", site="A")) or (fraction("Pb", site="A") + fraction("Bi", site="A") >= 0.05 and fraction("Pb", site="A") + fraction("Bi", site="A") <= 0.30) or (fraction("La", site="A") + fraction("Nd", site="A") + fraction("Pr", site="A") + fraction("Sm", site="A") + fraction("Ce", site="A") + fraction("Gd", site="A")) >= 0.20) and not has_element("Ni", site="B") |
| Co + (Ru or Ir) with strong lanthanide /Ca/(±Bi/Pb) A-anchor (anchor_A >= 0.50); exclude Ce and Ba; forbid Bi+Pb co-occupancy and Bi+Pb dominance (>0.50). | has_element("Co", site="B") and (has_element("Ru", site="B") or has_element("Ir", site="B")) and (fraction("La", site="A") + fraction("Nd", site="A") + fraction("Pr", site="A") + fraction("Y", site="A") + fraction("Gd", site="A") + fraction("Sm", site="A") + fraction("Bi", site="A") + fraction("Pb", site="A")) >= 0.5 and not has_element("Ce", site="A") and not has_element("Ba", site="A") and not (has_element("Pb", site="A") and has_element("Bi", site="A")) and (fraction("Bi", site="A") + fraction("Pb", site="A")) <= 0.5 |
| Sr/Y-gated Co–Ir synergy: Co and Ir co-present on B with Sr or Y on A and no Ce on A enriches for low OER_OP. | has_element("Co", site="B") and has_element("Ir", site="B") and (has_element("Sr", site="A") or has_element("Y", site="A")) and not has_element("Ce", site="A") |
| Co–Ru narrow-pocket: Co-dominant with substantial Ru share (>=10%) and capped Ir (<=25%); require A-site steric support or compact lanthanide pairing. | fraction("Co", site="B") >= 0.40 and fraction("Ru", site="B") >= 0.10 and fraction("Ir", site="B") <= 0.25 and (has_element("Ca", site="A") or has_element("Sr", site="A") or (min_property("atomic_number", "A") >= 57 and (mean_property("atomic_number", "A") - min_property("atomic_number", "A")) <= 4)) |

**Table S11.** Run level OER design performance and rule store statistics for the DSL-guided strategy.

| Run | Hit Rate | # of Rules | # of Verified Rules | # of Unverified Rules | Mean Precision | Mean Lift | # of Matched Data |
|---|---|---|---|---|---|---|---|
| 1 | 0.236 | 10 | 8 | 2 | 0.836 | 0.137 | **1102** |
| 2 | 0.123 | 10 | 8 | 2 | 0.685 | 0.073 | 612 |
| 3 | 0.211 | 10 | 6 | 4 | 0.805 | 0.043 | 575 |
| 4 | 0.072 | 10 | 9 | 1 | 0.617 | 0.067 | 543 |
| 5 | 0.292 | 10 | 9 | 1 | 0.775 | 0.155 | 598 |
| 6 | 0.174 | 9 | 8 | 1 | 0.888 | 0.078 | 248 |
| 7 | 0.281 | 10 | 9 | 1 | 0.851 | 0.130 | 547 |
| 8 | **0.302** | 10 | 9 | 1 | 0.838 | 0.210 | 796 |
| 9 | 0.223 | 10 | 7 | 3 | **0.905** | **0.250** | 115 |
| 10 | 0.216 | 10 | 8 | 2 | 0.784 | 0.102 | 689 |

**Table S12.** Comparison of the complexity and expressiveness of natural language rules generated by the DSL-guided and rule-only strategies.

| Method | Average Length of Rules | # of Mentions of the Word Within Rules (per Run) | | |
|---|---|---|---|---|
| | | 'Surface' | 'If' | 'Require' |
| DSL-guided | 530 | 0.1 | 0.6 | 0.8 |
| Rule-only | 1281 | 16.7 | 10.1 | 10.3 |

**Table S13.** The Gibbs free energy correction values for adsorbates and gaseous molecules. All units are in eV.

| Adsorbates | ZPE | $\int C_p dT$ | -TS | Molecules | ZPE | $\int C_p dT$ | -TS |
|---|---|---|---|---|---|---|---|
| O* | 0.08 | 0.03 | -0.06 | $H_2$ | 0.28 | 0.09 | -0.41 |
| OH* | 0.37 | 0.04 | -0.07 | $H_2O$ | 0.57 | 0.10 | -0.67 |
| OOH* | 0.44 | 0.05 | -0.09 | | | | |